\documentclass[nofootinbib,aps,11pt]{revtex4-1}
\usepackage{graphicx}
\usepackage{subfigure} 
\usepackage{hyperref}
\usepackage{cancel}
\usepackage{amssymb}
\usepackage{textcomp}
\usepackage{amsmath}
\usepackage{bm}
\usepackage{times}
\usepackage{epsfig}
\usepackage{color}
\usepackage{mathrsfs}
\usepackage{booktabs}
\usepackage{makecell}
\newcommand{\eff}{{\rm eff}}
\newcommand{\SI}{{\rm SI}}

\newcommand{\MeV}{{\rm MeV}}
\newcommand{\GeV}{{\rm GeV}}

\newcommand{\SM}{{\rm SM}}
\newcommand{\eq}{{\rm eq}}
\newcommand{\cm}{{\rm cm}}

\begin{document}
	\title{\Large $Z^\prime$ Portal Dark Matter with Observable $\Delta N_{\eff}$}
	\bigskip
	\author{Ang Liu$^1$}
	\email{AL@jnxy.edu.cn}
	\author{Zhi-Long Han$^2$}
	\email{sps\_hanzl@ujn.edu.cn}
	\author{Fei Huang$^{2,3}$}
	\email{sps\_huangf@ujn.edu.cn}
	\affiliation{$^1$School of Physical Science and Electronic Engineering, Jining University, Shandong 273155, China}		
	\affiliation{$^2$School of Physics and Technology, University of Jinan, Jinan, Shandong 250022, China}
	\affiliation{$^3$State Key Laboratory of Dark Matter Physics, School of Physics and Astronomy, Shanghai Jiao Tong University, Shanghai 200240, China}	
	\date{\today}
	\begin{abstract} 
	In the conventional $Z^\prime$ portal dark matter scenario, the prediction of detectable dark matter $\chi$ typically relies on the collider sensitivities of $Z^\prime$ and direct detection, where the Majorana type right-handed neutrinos  are usually assumed. However, if the right-handed  neutrinos $\nu_R$ are Dirac type, they will contribute to the additional effective number of relativistic species $\Delta N_{\eff}$, which brings different detectable predictions for $Z^\prime$ portal dark matter. In light of the great improvement of $\Delta N_{\eff}$ for the upcoming experiments, we investigate the $Z^\prime$ portal dark matter with Dirac type $\nu_R$. Under the $U(1)_{B-L}$ symmetry, this model  includes $\nu_R$ with $U(1)_{B-L}$ charge $Q_{\nu_R}=-1$  and $\chi$ with arbitrary $Q_\chi$ beyond the SM. Based on the relation in the production of $\chi$ and $\nu_R$, both the WIMP and FIMP dark matter through the $Z^\prime$ portal scenario are considered. We perform a comprehensive exploration of the viable parameter space under the constraints from $\Delta N_{\eff}$ induced by thermal and non-thermal $\nu_R$,  perturbative limit, dark matter direct and  indirect detection, and  collider searches of $Z^\prime$.	
	\end{abstract}
	
	\maketitle

\section{Introduction}
The mainstream opinion holds that  the Standard Model (SM) does not contain any viable dark matter candidate and predicts massless neutrinos \footnote{The recent study proposes SM neutrinos as potential dark matter candidates\cite{Cline:2026tkp} via enhancing their density.}. Currently, the cosmological and astrophysical observations  provide robust evidence for the existence of dark matter (DM)\cite{Cirelli:2024ssz}. Meanwhile, the neutrino oscillation observations indicate that neutrinos have sub-eV masses\cite{Super-Kamiokande:1998kpq,SNO:2002tuh,DayaBay:2012fng,RENO:2012mkc}. However, the fundamental natures of DM and neutrinos remain unclear. DM may consist of particles\cite{Bertone:2004pz}, in which the weakly interacting massive particle (WIMP)\cite{Steigman:1984ac,Arcadi:2017kky} and feebly interacting massive particle (FIMP)\cite{Hall:2009bx,Bernal:2017kxu} are the popular DM candidates.  The experiments searching for neutrinoless double-beta decay \cite{Dolinski:2019nrj} do not yield a definitive signal for Majorana neutrinos. Another alternative is Dirac neutrinos, which considerably contribute to the effective number of relativistic neutrino species $N_{\eff}$ via new interactions in new physics models \cite{Heeck:2014zfa,Abazajian:2019oqj,EscuderoAbenza:2020cmq,Luo:2020sho,Luo:2020fdt,Adshead:2022ovo,Babu:2022ikf,Esseili:2023ldf,Ghosh:2024cxi}. The induced cosmological constraints impose stringent limits on these models incorporating dark matter \cite{Biswas:2021kio,Borah:2022obi,Ghosh:2022fws,Borah:2022enh,Biswas:2022vkq,Ghosh:2023ocl,Borah:2023dhk,Mahapatra:2023oyh,Borah:2024gql,Borah:2024twm,Borboruah:2024lli,Oliveira:2025kfg,Abdelrahim:2025fiz,Okawa:2025sam,Das:2026xjq,Adhikary:2026ajn}.

The gauged $B-L$ extension of the SM is a popular new physics framework \cite{Davidson:1978pm,Mohapatra:1980qe,Marshak:1979fm,Masiero:1982fi,Mohapatra:1982xz,Buchmuller:1991ce,Wang:2025fya}, in which any anomalies could be resolved by introducing three right-handed neutrinos $\nu_R$, then producing the tiny neutrino masses via the seesaw mechanism \cite{Minkowski:1977sc,Mohapatra:1979ia,Schechter:1980gr,Schechter:1981cv}. Continuing to  extend the dark sector enables connections with DM, the cosmological observations of $N_{\eff}$ induced by Dirac neutrinos $\nu_R$ could achieve the detection of DM and the $Z^\prime$ boson \cite{Han:2018zcn,FileviezPerez:2019cyn,Mahanta:2021plx,Berbig:2022nre,Das:2023yhv,Borah:2025fkd,Herbermann:2025uqz,Ma:2026tyk}. Focus on the WIMP dark matter in the resonant scenario, the minimal extension involving only Dirac fermionic DM $\chi$ investigates the  constraints from $\Delta N_{\eff}$ on the $U(1)_{B-L}$ charge $Q_\chi$ of DM \cite{Han:2020oet}.  Besides the resonance scenario \cite{Duerr:2015wfa,Liu:2024esf}, DM could be produced via the secluded and FIMP scenarios \cite{Mohapatra:2019ysk} in this minimal $U(1)_{B-L}$ model, which has not been considered with Dirac neutrino $\nu_R$ and viable region under $\Delta N_{\eff}$.

The previous Planck data gives $\Delta N_{\eff}=N_{\eff}-N_{\eff}^{\SM}\lesssim0.285$ \cite{Planck:2018vyg}, where the contribution from the SM is $N_{\eff}^{\SM}=3.045$ \cite{Mangano:2005cc,Grohs:2015tfy,deSalas:2016ztq}. The recent results from DESI 2024 data give a slightly weaker bound $\Delta N_{\eff}\lesssim0.4$ \cite{DESI:2024mwx}, and P-ACT \cite{AtacamaCosmologyTelescope:2025blo,AtacamaCosmologyTelescope:2025nti} combined with Planck data pushes the bound down to $\Delta N_{\eff}\lesssim0.17$ at $95\%$ C.L.. The future CMB-S4 experiment will achieve higher sensitivity of $\Delta N_{\eff}\lesssim0.06$ \cite{Abazajian:2019eic}, and CMB-HD further narrows the range to $\Delta N_{\eff}\lesssim0.027$ \cite{CMB-HD:2022bsz}. These more stringent constraints will impose stronger limits on the charge $Q_\chi$ of DM. Thus, further investigating the impact of $\Delta N_{\rm eff}$ constraints on various DM scenarios is highly valuable.

In this work,  we construct the minimal $U(1)_{B-L}$ model that includes Dirac type right-handed  neutrinos $\nu_R$ and dark matter $\chi$. Depending on the production mechanism of $\chi$, it can be either a WIMP or a FIMP candidate. In the WIMP scenario with resonance and secluded cases, $\chi$ has the $U(1)_{B-L}$ charge $Q_\chi$. The value of $|Q_\chi|$  required to reproduce the observed dark matter relic density is typically much larger than $|Q_{f}|$, which is the $U(1)_{B-L}$ charge of SM fermions. Compared with the conventional generation of $\Delta N_{\eff}$ through $\nu_R \bar{\nu}_R \to f \bar{f}$ \cite{Heeck:2014zfa,Abazajian:2019oqj}, we systematically quantify the additional contribution from the newly introduced $\nu_R \bar{\nu}_R \to\chi \bar{\chi}$ and obtain the total experimental constraints of $\Delta N_{\eff}$. By combining the direct detection constraints, indirect detection constraints, the perturbative limit, and the collider searches for the $Z^\prime$ boson, we identify the viable parameter space compatible with the observed dark matter abundance. In the FIMP scenario, we investigate the production of Dirac neutrinos in the non-thermal region of DM. The inducing $\Delta N_{\eff}$ constraints, together with the collider signals of $Z^\prime$, are used to constrain the parameter space for FIMP dark matter.

The structure of this paper is organized as follows. In Section \ref{SEC:TM}, we provide a brief introduction to the theoretical  model employed in our study. We investigate the allowed parameter space of WIMP dark matter in Section \ref{SEC:WIMP} and that of FIMP dark matter in Section \ref{SEC:FIMP} under various constraints. Finally, we summarize the results in Section \ref{SEC:CL}.

\section{The model}\label{SEC:TM}

In the minimal $U(1)_{B-L}$ model, all anomalies are canceled by introducing three right-handed neutrinos $\nu_R$, which  carry  $U(1)_{B-L}$ charges $Q_{\nu_R}=-1$ \footnote{An alternative $U(1)_{B-L}$ charge assignment for the three right-handed neutrinos is $Q_{\nu_R}=(5,-4,-4)$ \cite{Montero:2007cd,Prajapati:2024wuu}.}.  The Dirac nature of $\nu_R$ is protected by the unbroken $B-L$ symmetry \cite{Heeck:2014zfa}. The neutrinos acquire masses through the Higgs mechanism 
\begin{equation}
	\mathcal{L}_\nu=-y\bar{L}\tilde{H}\nu_R+\text{h.c.}
\end{equation}
To produce sub-eV neutrino masses, the Yukawa coupling $y\lesssim10^{-11}$ is required. Contributions of this Yukawa interaction to $\Delta N_{\rm eff}$ is calculated as $\Delta N_{\rm eff}\approx 7.5\times10^{-12}$ \cite{Luo:2020fdt}, thus it is negligible.

The vector-like Dirac dark matter $\chi$ carries  an arbitrary $U(1)_{B-L}$ charge $Q_\chi$,  which does not affect the  anomaly cancellation.  To ensure the stability,  the dark matter  $\chi$ is CP-odd under an additional $Z_2$ symmetry, whereas all other particles are CP-even. For the unbroken $B-L$ symmetry, the new gauge boson $Z'$ is induced via the Stueckelberg mechanism \cite{Feldman:2006wb}. Under this configuration, the relevant Lagrangian can be written as
\begin{eqnarray}
	\mathcal{L}\supset -Z^\prime_{\mu}Q_{\chi}g^\prime\bar{\chi}\gamma^\mu\chi-Z^\prime_{\mu}Q_fg^{\prime}\bar{f}\gamma^\mu f
\end{eqnarray}
where $Q_f=\{Q_{\nu_R}, Q_{l}, Q_{q}\}$, the $U(1)_{B-L}$ charge  $Q_{l}=-1$ for SM leptons, and $Q_{q}=1/3$ for  quarks.  The free parameters are $\{m_\chi,m_{Z^\prime},g^\prime,Q_\chi\}$ in our work. Moreover, for simplicity, the notation $r_{Z^\prime}=m_{Z^\prime}/m_\chi$ is employed in the text.

\section{WIMP scenario}\label{SEC:WIMP}

\subsection{Relic density and $\Delta N_{\eff}$}\label{W-rd}

In the WIMP scenario, depending on the relative magnitudes of $m_{Z^\prime}$ and $m_\chi$, the relic density of DM is obtained via either the  $Z^\prime$ mediated resonance process $\chi\bar{\chi}\to f \bar{f}$ or the $\chi$ mediated secluded process $\chi\bar{\chi}\to Z^\prime Z^\prime$. The corresponding Boltzmann equation is
\begin{eqnarray}\label{Eqn:yc-w}
	\frac{dY}{dx_\chi} = -\frac{s}{\mathcal{H}x_\chi} \langle \sigma v\rangle_{\chi\bar{\chi}\to f\bar{f},Z^\prime Z^\prime}\Big(Y_{\chi}^2-(Y_{\chi}^{\eq})^2\Big),
\end{eqnarray}
where $x_\chi=m_{\chi}/T$, the entropy density $s=2\pi^2 (g_s(T)+21/4) T^3/45$.  The Hubble expansion rate is defined as $\mathcal{H}=\sqrt{4\pi^3(g_*(T)+21/4)/45} T^2/m_\text{pl}$ with the Planck mass $m_\text{pl}=1.22\times10^{19}~\GeV$. The factor $21/4$ comes from the contribution of three generations $\nu_R$. If dark matter $\chi$ decouples after $\nu_R$, this factor could be omitted. For clarity, we  summarize in Equation~\eqref{Eqn:Hubble}  whether this factor is neglected in different cases:
\begin{eqnarray}\label{Eqn:Hubble} 
	\begin{cases}
		\text{Thermal DM with}~T_{\rm dec}^{\chi}>T_{\rm dec}^{\nu_R}, & \mathcal{H}~\text{has}~21/4,\\
		\text{Thermal}~\nu_R, & \mathcal{H}~{\rm has}~21/4,\\
		\text{Thermal DM with}~T_{\rm dec}^{\chi}<T_{\rm dec}^{\nu_R}, & \mathcal{H}~\text{has no}~21/4,\\
		\text{Non-thermal DM}, & \mathcal{H}~\text{has no}~21/4,\\
		\text{Non-thermal} ~\nu_R, & \mathcal{H}~\text{has no}~21/4.\\
	\end{cases}
	\end{eqnarray}
$g_s(T)$ and $g_\star(T)$  are the number of relativistic degrees of freedom for the entropy density and energy density in SM, respectively, whose values are numerically calculated by micrOMEGAs \cite{Belanger:2013oya,Alguero:2023zol}. It should be stated here that in the more accurate Boltzmann equation, a factor of $d \log g_s(T)/ (3~d \log T)$ should be included on the right‑hand side of Equation \eqref{Eqn:yc-w}. We show in the Figure~\ref{FIG:fig8} (b) of Appendix~\ref{SEC:AEDT} the evolution of this factor with the temperature $T$, which remains much smaller than 1 within the decoupling temperature range of dark matter studied in this work. Therefore, we neglect this contribution for simplicity. 

The abundances of the non-relativistic $\chi$ and the relativistic  $f$  at the thermal equilibrium  are expressed as
\begin{eqnarray}
	Y_{\chi}^{\eq}=\frac{45m_{\chi}^2}{2\pi^4g_sT^2}\mathcal{K}_2(x_\chi),~Y_{f}^{\eq}=\frac{135\zeta(3)}{4\pi^4g_s}.
\end{eqnarray}

For the production of dark matter, the thermally averaged cross section of dark matter annihilation could be calculated by the numerical integration  over the  squared center-of-mass energy $s^\prime$,
	\begin{eqnarray}\label{Eqn:cfz-full}
		\langle \sigma v\rangle_{\chi\bar{\chi}\to f\bar{f},Z^\prime Z^\prime} &=& \frac{m_\chi}{16 \pi^4 x_\chi ({n_\chi^\eq})^2}	\int_{4 m_{\chi}^2}^\infty 2 (s^{\prime}- 4 m_{\chi}^2) \sigma(s^\prime)_{\chi\bar{\chi}\to f\bar{f},Z^\prime Z^\prime} \sqrt{s^\prime} \mathcal{K}_1 \left(\frac{x_\chi \sqrt{s^\prime}}{m_{\chi}}\right) ds^\prime
	\end{eqnarray}
where $\mathcal{K}_{1,2}$ are the modified Bessel functions of the second kind and $n_{i}^\eq=Y_{i}^\eq\times s$. The cross sections $\sigma(s^\prime)_{\chi\bar{\chi}\to f\bar{f},Z^\prime Z^\prime}$  of various processes can be found in Appendix~\ref{SEC:ACS}.	

Similar to the freeze-out mechanism of WIMP DM, when $\nu_R$ interacts strongly with the thermal bath, it attains thermal equilibrium and subsequently decouples from the bath at a decoupling temperature $T_{\rm dec}^{\nu_R}$. Its contribution to $\Delta N_{\eff}$ can be parameterized as \cite{Abazajian:2019oqj}
\begin{eqnarray}\label{Eqn:neff-w}
	\Delta N_\text{eff} &\simeq 0.047\times3\times\left(\frac{106.75}{g_\star (T_\text{dec}^{\nu_R})}\right)^{4/3}.
\end{eqnarray}
The detailed derivation  of this equation is presented in Appendix~\ref{SEC:ANEFF}. Here $g_\star(T)$ involves only the SM particles and has the maximum value 106.75 above the electroweak scale. The determination of $T_{\rm dec}^{\nu_R}$ relies on the relation $\tilde{\Gamma}_{\nu_R}(T_{\rm dec}^{\nu_R})=H(T_{\rm dec}^{\nu_R})$, where the reaction rate of the primary process related to $\nu_R$ meets  $\tilde{\Gamma}_{\nu_R}=\tilde{\Gamma}_{\nu_R\bar{\nu}_R\to f\bar{f}}+\tilde{\Gamma}_{\nu_R\bar{\nu}_R\to \chi\bar{\chi}}$. Suppressed by ${g^\prime}^4$, the $\nu_R \nu_R\to Z^\prime Z^\prime$ process is neglected in our analysis. The first term needs to be evaluated through a full integration \cite{Barger:2003zh}
	\begin{eqnarray} \label{Eqn:vrf-full}
			\tilde{\Gamma}_{\nu_R\bar{\nu}_R\to f\bar{f}}&=& \frac{n_{\nu_R}^\eq}{2} \langle\sigma v \rangle_{\nu_R \bar{\nu}_R \to f\bar{f}}\\ \nonumber
			&=& \frac{2}{n_{\nu_R}^\eq}  \int \frac{d^3 \vec{p}}{(2\pi)^3} \frac{d^3 \vec{k}}{(2\pi)^3} f_\nu (p) f_\nu (k) v_\mathrm{M} \sigma(s^\prime)_{\nu_R\bar{\nu}_R\to f\bar{f}},
	\end{eqnarray}
where the Fermi-Dirac distribution $f_\nu (k) = (e^{k/T}+1)^{-1}$,  the M{\o}ller velocity $v_\mathrm{M}=1-\cos \theta$ and  $s^\prime = 2 p k (1-\cos\theta)$ with $\theta$~being the angle between the two colliding $\nu_R$ particles. The expression for the cross section $\sigma(s^\prime)_{\nu_R\bar{\nu}_R\to f\bar{f}}$  is given in  Appendix~\ref{SEC:ACS}. The second term is expressed as
\begin{eqnarray}
	\tilde{\Gamma}_{\nu_R\bar{\nu}_R\to \chi\bar{\chi}} = \frac{(n_{\chi}^\eq)^2}{n_{\nu_R}^\eq} \langle\sigma v_{\chi\bar{\chi}\to \nu_R\bar{\nu}_R} \rangle,
\end{eqnarray}
In practice, if DM decouples prior to $\nu_R$, the contribution of this process does not need to be computed.

For simplicity, the thermally averaged cross sections for various processes can be approximated as simpler analytical forms under certain conditions. For example, the thermal average cross section of $\chi\bar{\chi}\to f \bar{f}$  with  on-shell ${Z^\prime}$ could be approximated as \cite{Nath:2021uqb}
\begin{eqnarray}\label{Eqn:cf-on}
	\langle\sigma v \rangle^{\rm on}_{\chi\bar{\chi}\to f\bar{f}} \simeq \frac{3\pi^2 m_{Z^\prime}^2}{2 m_{\chi}^5}\times \frac{x_\chi~\mathcal{K}_1 \left(\frac{x_\chi \times m_{Z^\prime}}{m_\chi}\right)}{\Big(\mathcal{K}_2(x_\chi)\Big)^2} \times \frac{\Gamma_{Z^\prime\to \chi\bar{\chi}} \Gamma_{Z^\prime\to f\bar{f}}}{\Gamma_{Z^\prime}},
\end{eqnarray}
For off-shell ${Z^\prime}$, the thermal average cross section of $\chi\bar{\chi}\to f \bar{f}$ is estimated as \cite{Okada:2020cue}
\begin{eqnarray}\label{Eqn:cf-off}
	\langle\sigma v \rangle^{\rm off}_{\chi\bar{\chi}\to f\bar{f}} \simeq\frac{13 Q_\chi^2 Q_f^2 {g^\prime}^4 x_\chi^2}{192\pi m_{\chi}^2}.
\end{eqnarray}
where we have multiplied the final result by a factor of 2 for Dirac $\chi$.  When $m_\chi>m_{Z^\prime}$, the analytical expression of  $\langle\sigma v \rangle_{\chi \chi\to Z^\prime Z^\prime}$  is \cite{Mohapatra:2019ysk}
\begin{eqnarray}\label{Eqn:cz}
	\langle\sigma v\rangle_{\chi \chi\to Z^\prime Z^\prime}  \simeq  \frac{{Q_\chi}^4 {g^\prime}^4}{16\pi m_{\chi}^2} \left(1-\frac{m_{Z^\prime}^2}{m_{\chi}^{2}} \right)^{3/2}
	\left(1-\frac{m_{Z^\prime}^2}{2 m_{\chi}^{2}} \right)^{-2}. 
\end{eqnarray}
For the production of $\nu_R$,  the Equation~\eqref{Eqn:vrf-full}  could  be simplified to \cite{Heeck:2014zfa}
\begin{eqnarray}\label{Eqn:vrf}
	\tilde{\Gamma}_{\nu_R\bar{\nu}_R\to f\bar{f}} = \frac{N_c^f {g^\prime}^4 Q_f^2 T}{36 \pi^3 \zeta (3)}
	\times \begin{cases}
		\frac{\pi^4}{144}, & x_{Z^\prime} \lesssim \sqrt{\epsilon},\\
		\frac{1.15\pi}{8} \frac{m_{Z^\prime}}{\Gamma_{Z^\prime}} \frac{x_{Z^\prime}^3}{e^{x_{Z^\prime}} -1} , & \sqrt{\epsilon} \lesssim  x_{Z^\prime} \lesssim 14 \sqrt{\log \epsilon^{-1}},\\
		\frac{49\pi^8}{2700} x_{Z^\prime}^{-4}, & x_{Z^\prime} \gtrsim 14 \sqrt{\log \epsilon^{-1}},
	\end{cases}
\end{eqnarray}
where  $x_{Z^\prime}=m_{Z^\prime}/T$ and $\epsilon=\Gamma_{Z^\prime}/m_{Z^\prime}$. The assumptions made in these approximate results are summarized in Table \ref{Tab:Assumptions}.

\begin{table}
	\begin{center}\large
		\begin{tabular}{c |c | c } 
			\hline
			Process &  Scenario & Assumptions  \\ \hline
			Equation~\eqref{Eqn:cf-on} & \makecell{Thermal DM and  $\nu_R$ \\ (on-shell $Z^\prime$) }& Narrow width approximation with $x_\chi\gtrsim1$   \\ \hline
			Equation~\eqref{Eqn:cf-off} &Thermal DM and  $\nu_R$  & $m_\chi\gg m_{Z^\prime}$, $x_\chi\lesssim1$ and $m_f=0$ \\ \hline
			Equation~\eqref{Eqn:cz} &Thermal DM& \makecell{S-wave approximation with \\ the nonrelativistic limit } \\ \hline
			Equation~\eqref{Eqn:vrf} &Thermal/Non-thermal $\nu_R$ & Narrow width approximation  \\ \hline
			Equation~\eqref{Eqn:r-id} &Thermal DM & \makecell{S-wave approximation with \\ the nonrelativistic limit }  \\ \hline
			Equation~\eqref{Eqn:cf-fimp} &Non-thermal DM& \makecell{Narrow width approximation with \\ $x_\chi\ll1$ and $m_f=0$}   \\ \hline
		\end{tabular}
	\end{center}
	\caption{Summary of the assumptions in the approximate thermally averaged cross sections.
		\label{Tab:Assumptions}}
\end{table}

\begin{figure}
	\begin{center}
		\includegraphics[width=0.45\linewidth]{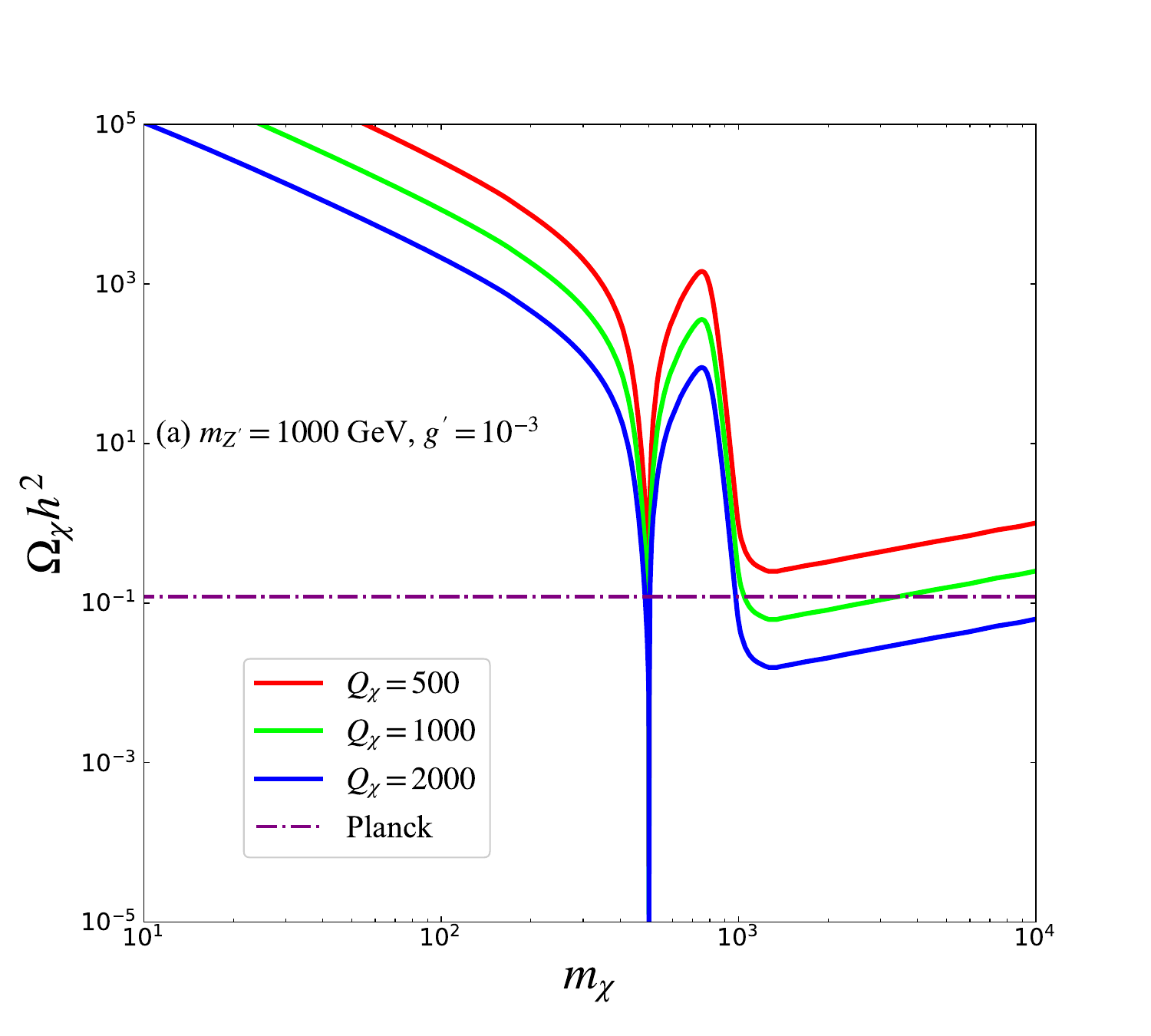}
		\includegraphics[width=0.45\linewidth]{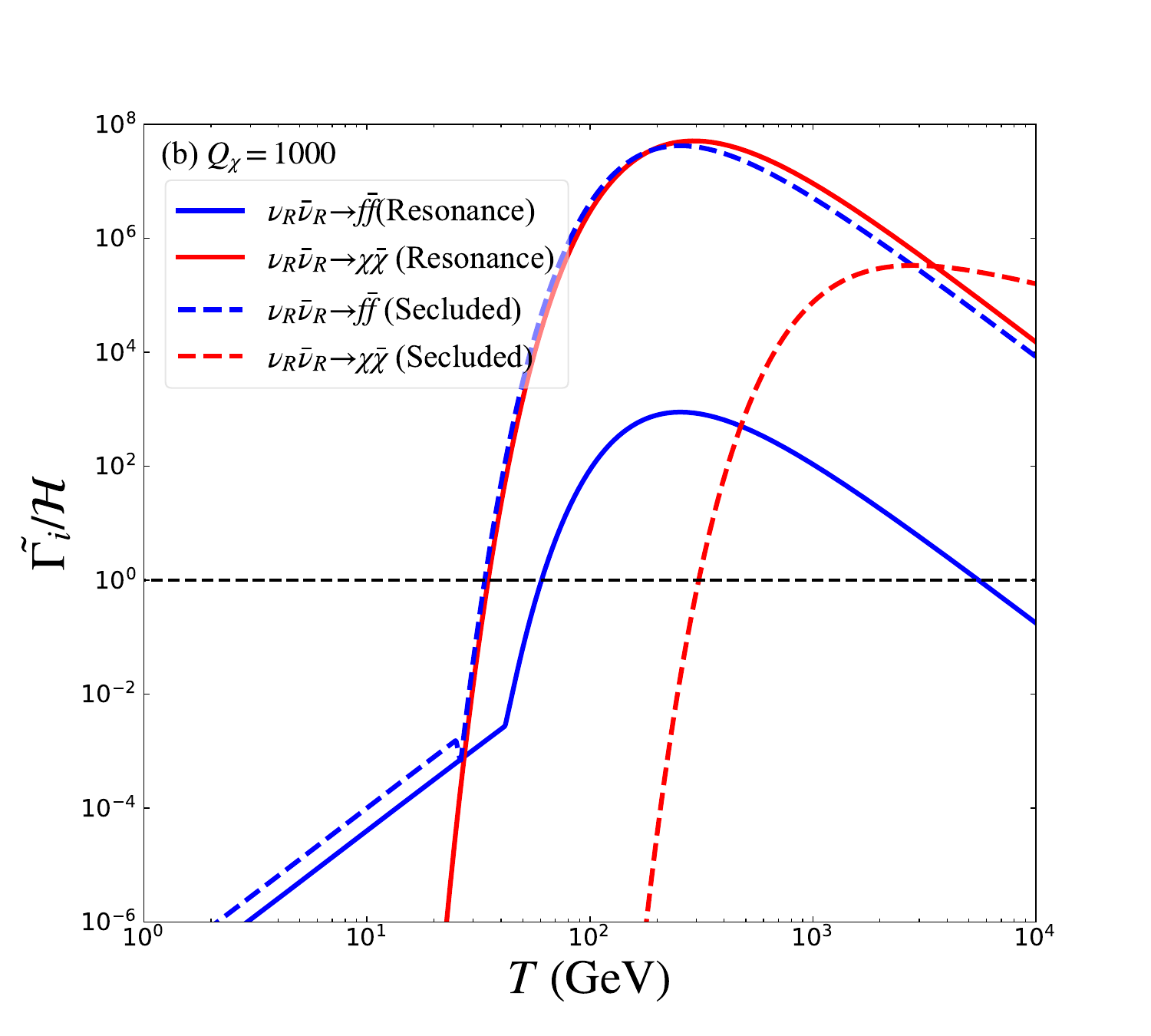}
		\includegraphics[width=0.45\linewidth]{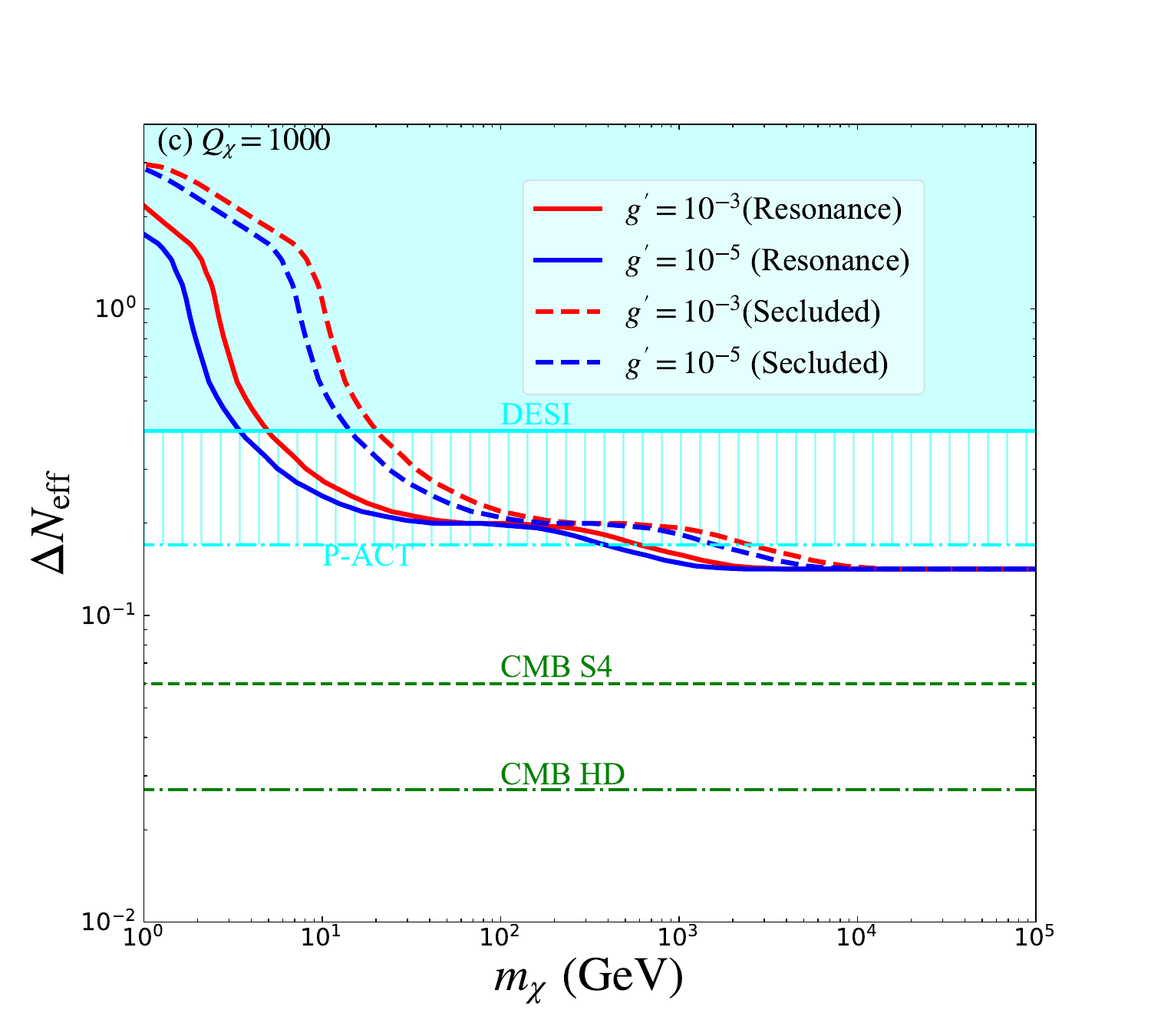}
	\end{center}
	\caption{Panel (a): The dependence of DM relic density in the WIMP scenario. Panel (b): Influence of different processes on the  decoupling temperature of $\nu_R$. Panel (c): The dependency of $\Delta N_{\eff}$ on $m_\chi$ in the resonance and secluded scenarios. In panel (a), the red, green, and blue curves correspond to $Q_\chi=500$, $Q_\chi=1000$, and $Q_\chi=2000$, respectively. The horizontal purple dot-dashed line denotes the observed dark matter relic density from the Planck experiment, i.e. $\Omega_{\chi}h^2=0.12$~\cite{Planck:2018vyg}. In panel (b),  $m_{Z^\prime}$ and $g^\prime$ are consistent with those in panel (a). Both the resonance benchmark point with $m_\chi=497$ GeV and the secluded one with  $m_\chi=3400$ GeV could produce the observed DM relic density. The red and blue solid lines represent the two distinct processes related to $\nu_R$ generation in the resonance scenario, while the dashed lines represent those in the secluded one. In panel (c), the solid and dashed lines represent the resonance and secluded scenarios, with the corresponding $r_{Z^\prime}$ being 2.001 and 0.5, respectively. The red and green colors denote two different $g^\prime=10^{-3},10^{-5}$. The cyan solid and dot-dashed lines represent the upper limits on $\Delta N_{\eff}$ given by DESI 2024 $\Delta N_{\eff}\simeq0.4$ \cite{DESI:2024mwx} and P-ACT $\Delta N_{\eff}\simeq0.17$ \cite{AtacamaCosmologyTelescope:2025blo,AtacamaCosmologyTelescope:2025nti}, respectively. The green dashed and dot-dashed horizontal lines are the future sensitivity of CMB-S4 $\Delta N_{\eff}\simeq0.06$ \cite{Abazajian:2019eic} and CMB-HD $\Delta N_{\eff}\simeq0.027$ \cite{CMB-HD:2022bsz}.
	}
	\label{FIG:fig1}
\end{figure}

Next, we select an appropriate benchmark point $(m_{Z^\prime}=1000~\GeV,~g^\prime=10^{-3})$  to illustrate the dependence of  DM relic density on  $m_\chi$ in Figure~\ref{FIG:fig1},  which can satisfy the strict collider constraints of $Z^\prime$. Solutions satisfying DM observations clearly emerge at the resonant  position $m_{Z^\prime}\simeq 2m_{\chi}$ and the secluded regions $m_{Z^\prime}\lesssim m_{\chi}$. In contrast, the secluded case requires a larger $Q_\chi$. Therefore, the subsequent WIMP study focuses on these two distinctive cases. Moreover, for convenience, we adopt the approximate condition that yields the observed DM relic abundance derived in the secluded scenario \cite{Mohapatra:2019ysk}, namely, $g^\prime\simeq1.6\times10^{-2}~\sqrt{m_{\chi}}/Q_\chi$, in the subsequent phenomenological discussion.

\subsubsection{Resonance case}

Under the premise that the DM relic density satisfies the observed value, we validate the impact of $\nu_R$-related processes on  its decoupling temperature in panel (b) of Figure~\ref{FIG:fig1}.  In the resonant benchmark, the reaction rates $\tilde{\Gamma}_{\nu_R\bar{\nu}_R\to \chi\bar{\chi}}$ and $\tilde{\Gamma}_{\nu_R\bar{\nu}_R\to f\bar{f}}$ are proportional to $\Gamma_{Z^\prime\to \chi\bar{\chi}} \Gamma_{Z^\prime\to \nu_R\bar{\nu_R}}/\Gamma_{Z^\prime}$ and $\Gamma_{Z^\prime\to f\bar{f}} \Gamma_{Z^\prime\to \nu_R\bar{\nu_R}} /\Gamma_{Z^\prime}$, respectively. Due to the large $Q_\chi$ under non-extreme resonance conditions for the benchmark, we have $\Gamma_{Z^\prime\to \chi\bar{\chi}}\gg\Gamma_{Z^\prime\to f\bar{f}}$.  Therefore, the magnitude of $\tilde{\Gamma}_{\nu_R\bar{\nu}_R\to \chi\bar{\chi}}$ far exceeds that of $\tilde{\Gamma}_{\nu_R\bar{\nu}_R\to f\bar{f}}$.  Moreover, for $\tilde{\Gamma}_{\nu_R\bar{\nu}_R\to \chi\bar{\chi}}$ to truly play a role in the decoupling of $\nu_R$, it must be satisfied that dark matter decouples after $\nu_R$, i.e., $T_{\rm dec}^{\chi}<T_{\rm dec}^{\nu_R}$. In this benchmark point,  $\nu_R$ decouples when $T\simeq36$ GeV, meanwhile dark matter is still in the thermal equilibrium as $T_{\rm dec}^{\chi}\simeq m_\chi/25\simeq20$ GeV for WIMP DM. Substituting the corresponding $g_*(T_\text{dec}^{\nu_R})\simeq91.5$ into Equation~\eqref{Eqn:neff-w} yields $\Delta N_{\eff}=1.74\times10^{-1}$ with three $\nu_R$. Of course, if within the regime of extreme resonance, $\Gamma_{Z^\prime\to \chi\bar{\chi}}$ is strongly suppressed by the phase space. Correspondingly, the influence of $\nu_R\bar{\nu}_R\to \chi\bar{\chi}$ could be disregarded.

In panel (c) of Figure~\ref{FIG:fig1}, we examine the  influence of $m_\chi$ on $\Delta N_{\eff}$ for fixed values of $g'$. Notably, here we do not insist on the dark matter matching the observed relic density. The compatible results will be discussed in detail later  in the subsections~\ref{W-neff}. With fixed mass ratio $r_{Z'}$, the increase in $m_\chi$ essentially reflects a rise in $m_{Z^\prime}$, which in turn elevates $T_{\rm dec}^{\nu_R}$ and causes $g_\star (T_\text{dec}^{\nu_R})$ to increase. As a result, Equation~\eqref{Eqn:neff-w} shows that $\Delta N_{\eff}$ drops continuously until it flattens out at a minimum value of 0.14. This trend is fully determined by $g_\star (T_\text{dec}^{\nu_R})$.  A reduction in $g^\prime$ leads to only a slight variation in $g_\star (T_\text{dec}^{\nu_R})$, and thus $\Delta N_{\eff}$ changes little accordingly. Therefore, the future precise measurement of $\Delta N_{\eff}$ is promising to reveal the nature of DM. In the resonance case, $\Delta N_{\eff}$ meets the DESI and P-ACT bounds for $m_\chi\gtrsim3$~GeV and $m_\chi\gtrsim400$ GeV, respectively. All these permitted parts will be tested by the future CMB-S4 and CMB-HD experiments. 

\subsubsection{Secluded case}

In the secluded benchmark of Figure~\ref{FIG:fig1} (b), since $\nu_R\bar{\nu}_R\to \chi\bar{\chi}$ is suppressed by the off-shell $Z^\prime$, $\nu_R\bar{\nu}_R\to f\bar{f}$ becomes the dominant process.  It should be noted that the result in Equation~\eqref{Eqn:cf-off} is only valid for $x_\chi\lesssim1$, we numerically solve the Equation~\eqref{Eqn:cfz-full} to obtain the evolution of this process over the full range.  The reaction rate of $\nu_R\bar{\nu}_R\to f\bar{f}$ is almost the same as that of the resonant $\nu_R\bar{\nu}_R\to \chi\bar{\chi}$, which are both essentially proportional to $\Gamma_{Z^\prime\to f\bar{f}}$.  More importantly, DM with larger mass decouples earlier, so the decoupling of $\nu_R$ is only affected by $\nu_R\bar{\nu}_R\to f\bar{f}$. In Figure~\ref{FIG:fig1} (c), owing to the influence of $r_{Z^\prime}=0.5$, these two $m_\chi$ thresholds of the DESI and P-ACT bounds shift to 16 GeV and 1500 GeV , respectively. Eventually, $\Delta N_{\eff}$ drops to its minimum value of 0.14 when $m_\chi\gtrsim10^4$ GeV, which can be probed by future the future CMB-S4 and CMB-HD experiments.

Overall, in the resonance scenario, whether $\nu_R\bar{\nu}_R\to \chi\bar{\chi}$ participating in the decoupling of $\nu_R$ depends on the relative magnitudes of $T_{\rm dec}^{\chi}$ and $T_{\rm dec}^{\nu_R}$. Both processes take effect  when $T_{\rm dec}^{\chi}<T_{\rm dec}^{\nu_R}$, otherwise, only $\nu_R\bar{\nu}_R\to f\bar{f}$ works. In the secluded case, DM does not affect the decoupling of $\nu_R$. Furthermore, the thermally produced $\Delta N_{\eff}$ decreases with increasing $m_\chi$, and eventually remains constant at 0.14 with three generations of $\nu_R$. If no clear excess of $\Delta N_{\eff}$ is observed in the future, the minimal $Z'$ portal dark matter with Dirac neutrino can be fully excluded.

\subsection{Direct detection of dark matter}\label{W-dd}

\begin{figure}
	\begin{center}
		\includegraphics[width=0.45\linewidth]{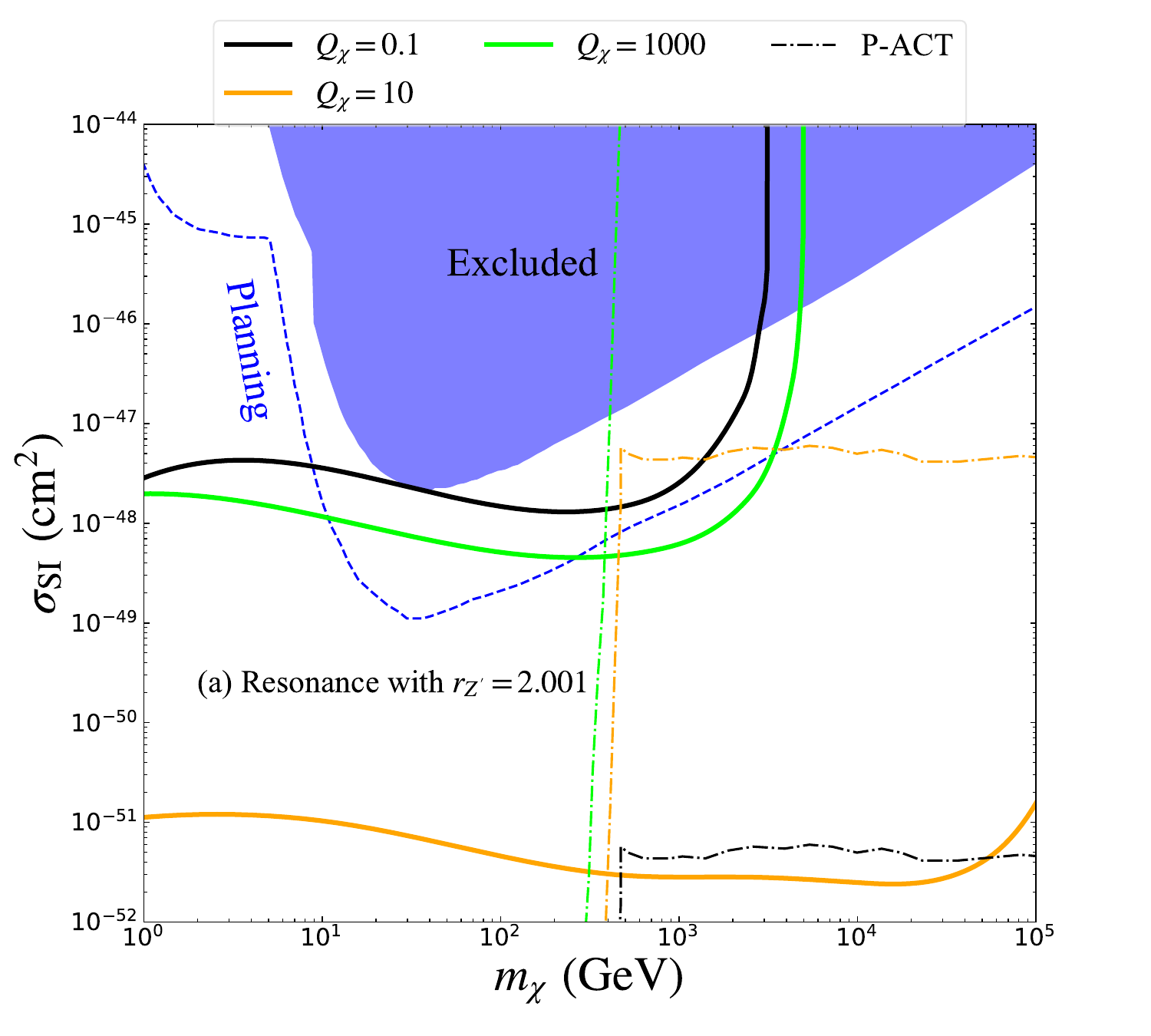}
		\includegraphics[width=0.45\linewidth]{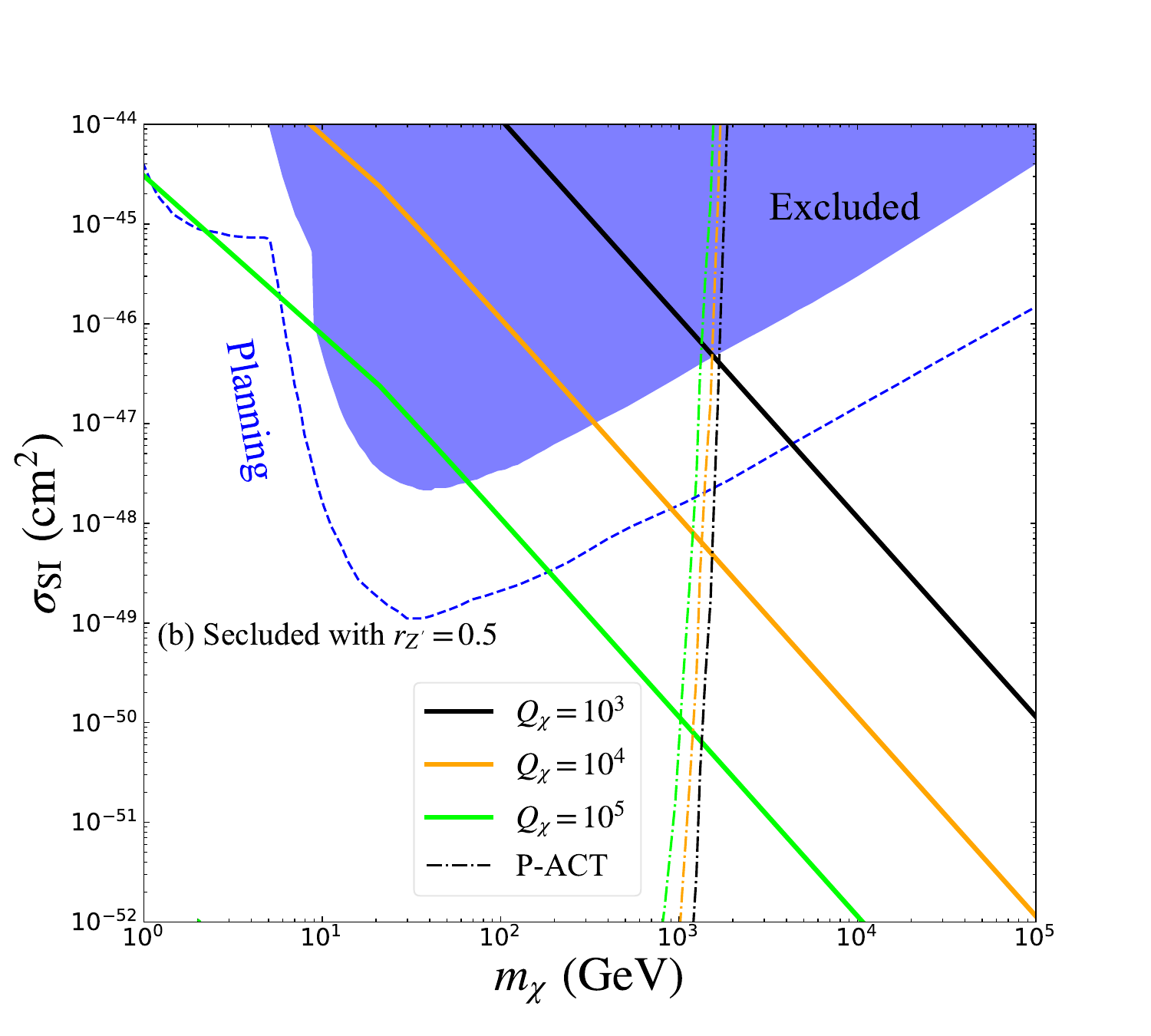}
	\end{center}
	\caption{Constraints from DM direct detection experiments and cosmological experiments related to $\Delta N_{\rm eff}$ in the resonance scenario (a) and secluded scenario (b). The blue region and dashed lines  denote the parameter space excluded by current direct detection experiments as well as the projected sensitivity  of upcoming experiments, respectively. 	
		The solid black, orange, and green curves represent three benchmark lines satisfying the DM observation. The corresponding dashed lines show the $\Delta N_{\eff}$ constraints from P-ACT expermiment \cite{AtacamaCosmologyTelescope:2025blo,AtacamaCosmologyTelescope:2025nti} with $Q_{\chi}=0.1$, $Q_{\chi}=10$ and $Q_{\chi}=1000$  in  the resonance scenario, with  $Q_{\chi}=10^3$, $Q_{\chi}=10^4$ and $Q_{\chi}=10^5$  in  the secluded scenario.   For each benchmark point, the region outside the corresponding dashed curve, namely, $\Delta N_{\eff}\gtrsim0.17$, is not permitted by P-ACT constraints.
	}
	\label{FIG:fig2}
\end{figure}

The direct detection experiments for DM are sensitive to the spin-independent DM–nucleon scattering cross section $\sigma_{\SI}$. At the canonical GeV-TeV mass range for WIMP DM, some current experiments DarkSide-50 \cite{DarkSide-50:2023fcw}, XENONnT \cite{XENON:2025vwd}, PandaX-4T \cite{PandaX:2024qfu}, and LZ \cite{LZ:2024zvo} have searched the parameter space  with  $\sigma_{\SI}\gtrsim\mathcal{O}(10^{-48})~\cm^2$. The absence of a detected DM signal  excludes this region, as indicated by the blue shaded area in Figure~\ref{FIG:fig2}. The upcoming experiment DarkSide-LowMass \cite{GlobalArgonDarkMatter:2022ppc}, SuperCDMS \cite{SuperCDMS:2016wui}, and LZ \cite{LZ:2015kxe} will achieve significantly improved sensitivity on $\sigma_{\SI}$, extending the reach downward by approximately one order of magnitude compared to current limits. The upcoming results are depicted as a blue dashed line in Figure~\ref{FIG:fig2}.

In this model, the DM-nucleon scattering process is mediated by $Z^\prime$, which can be calculated as
\begin{eqnarray}\label{Eqn:dd}
	\sigma_{\rm SI}=\frac{m_{\chi}^2 Q_\chi^2{g^\prime}^4 m_n^2}{\pi m_{Z^\prime}^4(m_{\chi}+m_n)^2},
\end{eqnarray}
where the mass of nucleons $m_n\simeq0.939$ GeV.  In Figure~\ref{FIG:fig2}, we select three benchmark points meeting the observed DM relic density for  the resonant and secluded cases to illustrate the influence from the direct  detection constraints as well as cosmological bounds of $\Delta N_{\eff}$ induced by $\nu_R$. 

\subsubsection{Resonance case}

In the resonance scenario, as shown in panel (a) of Figure~\ref{FIG:fig2},  as $Q_\chi$ increases from 0.1 to 1000, the corresponding $\sigma_{\rm SI}$  does not exhibit a monotonic trend.  We explain this by comparing the evolution of the corresponding benchmark in the $m_{Z^\prime}-g^\prime$ parameter space shown in Figure~\ref{FIG:fig4}.  Firstly, for any benchmark line that satisfies the DM relic density, the required $g^\prime$ is proportional to $m_\chi$ for relatively light DM.  As both increase together, when the effective couplings $g_\chi=g' Q_\chi$ reach the perturbativity limit of $\mathcal{O}(1)$, $\Gamma_{Z^\prime}$  undergoes a rapid rise, where the narrow-width approximation  is no longer suitable. Referring to the form of $\langle \sigma v\rangle$ that includes the Breit-Wigner propagator in Equation~\eqref{Eqn:r-id}, the contribution from the term $(4m_{\chi}^2-m_{Z^\prime}^2)^2$ is much smaller that of the $m_{Z'}^2\Gamma_{Z'}^2$ term, which leads to the cancellation of  the couplings $g'$ in the numerator and denominator. Therefore the benchmark line starts to become independent of the couplings $g'$, but the specific  cutoff position depends on $Q_\chi$.

Then, as $Q_\chi$ increases from 0.1 to 10, $\Gamma_{Z^\prime\to \chi\bar{\chi}}$ is smaller than $\Gamma_{Z^\prime\to f\bar{f}}$, and the narrow-width approximation of Equation~\eqref{Eqn:cf-on} simplifies to be proportional to $\Gamma_{Z^\prime\to\chi\bar{\chi}}$. So $g^\prime$ decreases proportionally with increasing $Q_\chi$.  However, as $Q_\chi$ continues to increase, $\Gamma_{Z^\prime\to \chi\bar{\chi}}$ has the dominant contribution. Equation~\eqref{Eqn:cf-on} is approximately proportional to $\Gamma_{Z^\prime\to f\bar{f}}$, thus the benchmark line no longer shows a significant downward trend with increasing $Q_\chi$. Under this principle, combined with Equation~\eqref{Eqn:dd}, one can obtain the rapidly decreasing $\sigma_{\rm SI}$  corresponding to $Q_\chi$ from 0.1 to 10. In addition, when $Q_\chi$ ranges from 10 to 100, $g^\prime$ changes very little, which causes the corresponding $\sigma_{\rm SI}$ to increase.

By comparing with the direct detection constraints, we find that the current limits cannot exclude the parameter space with $0.1\lesssim Q_\chi\lesssim1000$, except the maximum  mass cutoff region where the coupling $g_\chi$  exceeds the perturbative limit. The $Q_\chi$ outside this range is constrained by direct detection. The future direct detection experiments will be sensitive to $10~\GeV\lesssim m_\chi\lesssim1000~\GeV$ when $Q_\chi\sim0.1$ or 1000. However, the situation changes dramatically when the $\Delta N_{\eff}$ constraints are incorporated. The benchmark $Q_{\chi} = 0.1$ is entirely excluded by $\Delta N_{\eff}$ constraints, but $Q_{\chi}\gtrsim10$ could evade $\Delta N_{\eff}$ constraints within $m_{\chi} \gtrsim 400$ GeV.

\subsubsection{Secluded case}

In the secluded scenario, the dark matter direct detection constraints are displayed in panel (b) of Figure~\ref{FIG:fig2} with $r_{Z^\prime}=0.5$.  The calculated $\sigma_{\rm SI}$  survives only when $m_{\chi} \gtrsim 1500$ GeV and $\sigma_{\rm SI}\lesssim \mathcal{O}(10^{-46})~\cm^2$  under the combined constraints from the direct detection and $\Delta N_{\eff}$. Increasing the value of $Q_\chi$ leads to a smaller DM-nucleon scattering cross section for fixed value of $m_\chi$. The magnitude of the constrained $m_\chi$ is determined by $r_{Z^\prime}$, with a larger $r_{Z^\prime}$ yielding a larger $m_{\chi}$. Moreover, for $Q_\chi\sim\mathcal{O}(10^3)$, the upcoming direct detection experiments will be capable of probing $m_\chi$ above TeV within  $\mathcal{O}(10^{-48})~\cm^2\lesssim\sigma_{\rm SI}\lesssim\mathcal{O}(10^{-46})~\cm^2$.

In summary, compared to the direct detection constraints, the $\Delta N_{\eff}$ could impose stronger restrictions on  the WIMP scenario, which is particularly sensitive to $m_\chi$. Specifically, $m_\chi\lesssim400$ GeV is disallowed in the resonant case, while the secluded one rules out $m_\chi$ below TeV. 

\subsection{Indirect detection of dark matter}\label{W-id}

The present annihilation cross section $\langle \sigma v\rangle$ of DM into the SM particles can be detected by the indirect detection experiments. 

\subsubsection{Resonance case}

\begin{figure}
	\begin{center}
		\includegraphics[width=0.45\linewidth]{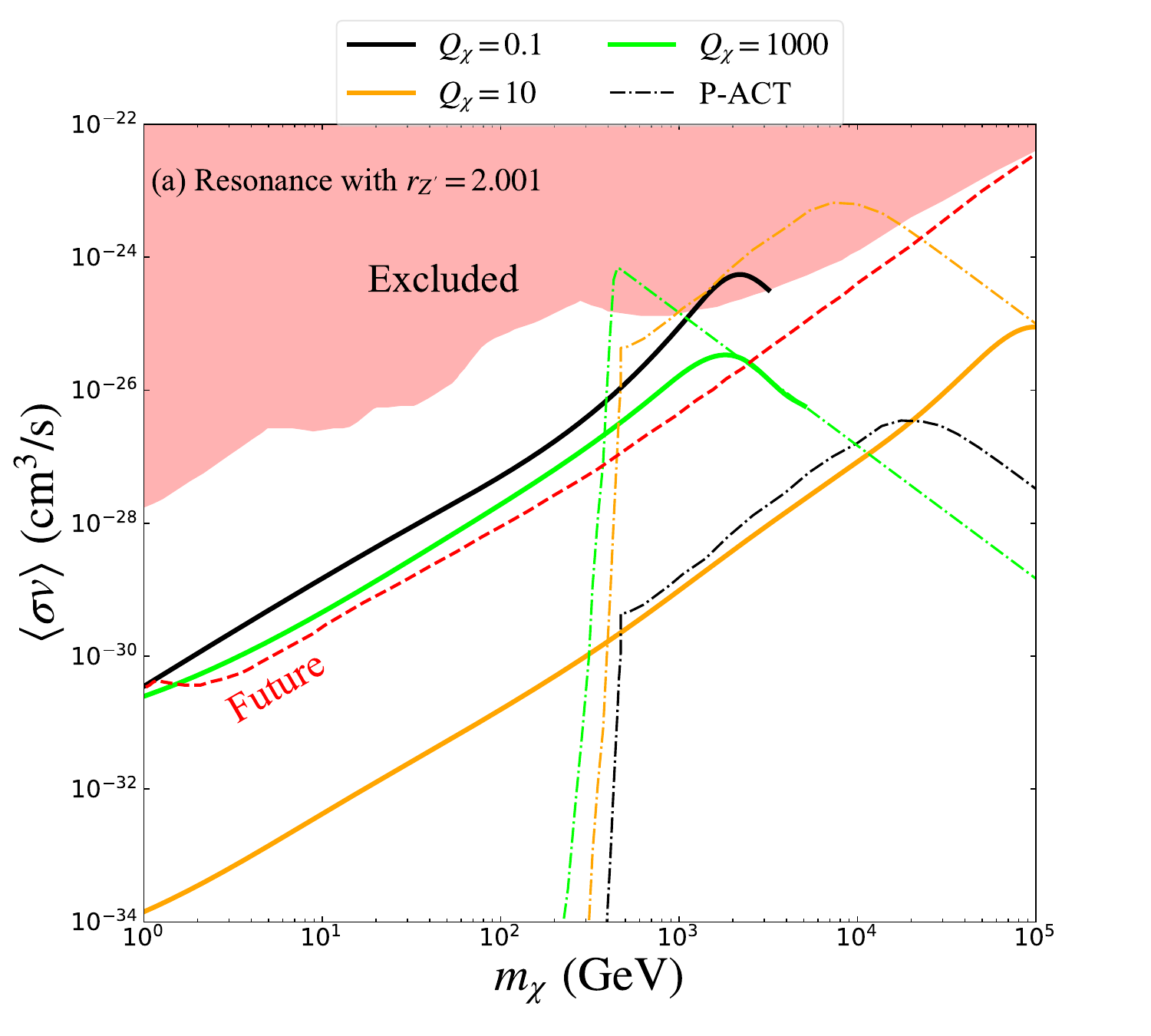}
		\includegraphics[width=0.45\linewidth]{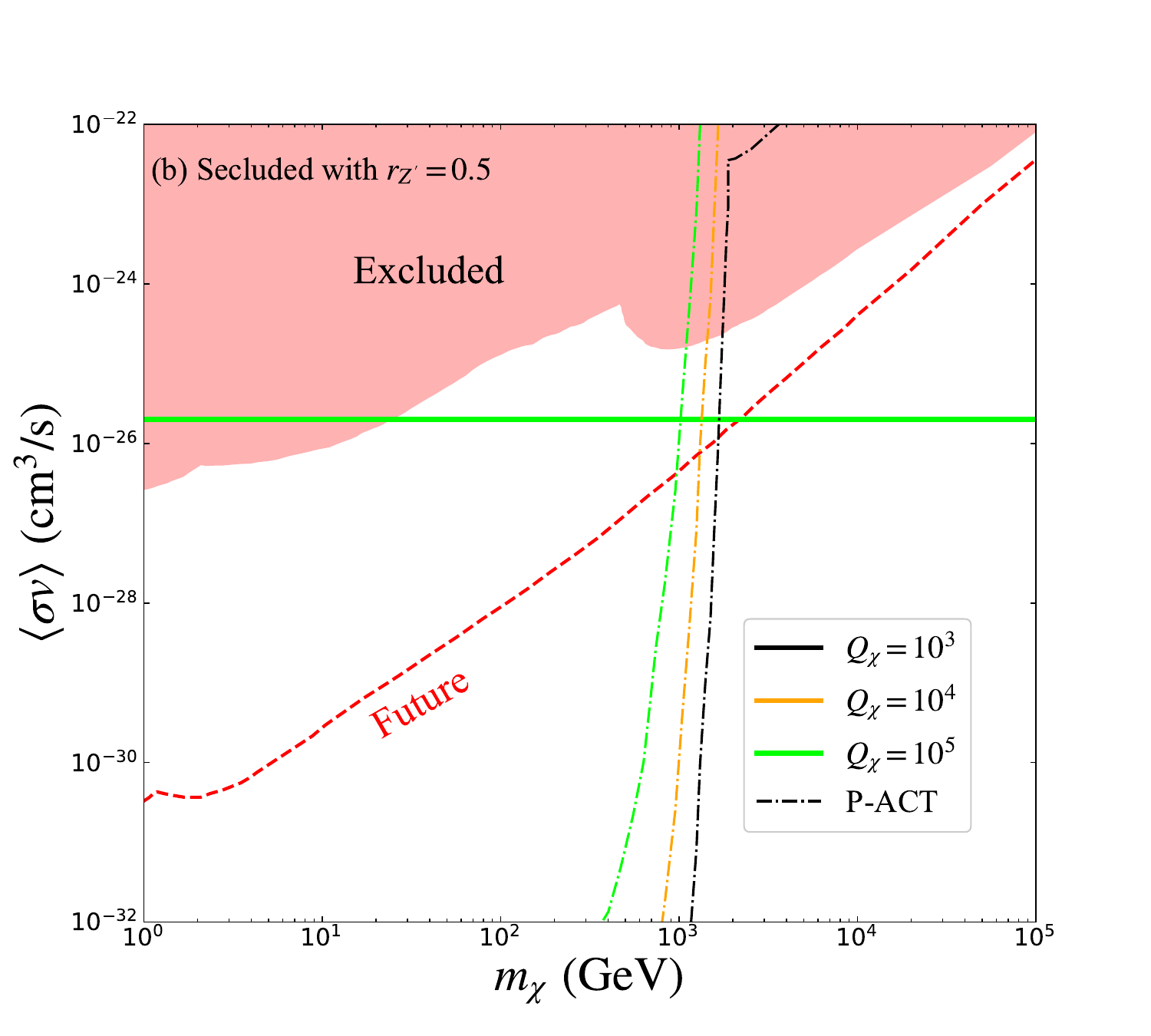}
	\end{center}
	\caption{The DM indirect detection and $\Delta N_{\rm eff}$ constraints in the resonance scenario (a) and the secluded scenario (b). In both panels, the red region is excluded by the current indirect detection experiments. The future experiments are expected to probe the parameter space enclosed by the red dashed curve. The benchmark curves share identical legends with those  in Figure~\ref{FIG:fig2}.
	}
	\label{FIG:fig3}
\end{figure}

In the resonance scenario, the present $\langle \sigma v\rangle$ is numerically calculated through 
\begin{eqnarray}\label{Eqn:r-id}
\langle \sigma v\rangle_{\chi\bar{\chi}\rightarrow f\bar{f}} &=& \frac{N_C^f Q^2_\chi Q^2_f {g^\prime}^4}{2\pi}\sqrt{1-\frac{m^2_f}{m^2_\chi}}\frac{2m^2_\chi+m^2_f}{(4m^2_\chi-m^2_{Z^\prime})^2+m^2_{Z^\prime}\Gamma^2_{Z^\prime}}.
\end{eqnarray}
Within this model, due to the setting of $Q_f$, the branching ratio into lepton final states is larger than that into quark final states. Therefore, we select the electron final state in Figure~\ref{FIG:fig3} to illustrate the indirect detection constraints.

For the resonance scenario in panel (a) of Figure~\ref{FIG:fig3}, the existing constraints with $m_{\chi}\lesssim5$ GeV come from experiments involving XMM-NEWTON $X$-rays \cite{Cirelli:2023tnx} and CMB (s-wave) \cite{Lopez-Honorez:2013cua,Slatyer:2015jla} observations.  While the results for larger $m_{\chi}$ are taken from literatures \cite{Leane:2018kjk,Dutta:2022wdi}, which are the convolutions of the bounds from AMS positron \cite{AMS:2014xys,AMS:2019rhg}, Fermi-LAT dwarfs \cite{Fermi-LAT:2016uux} and H.E.S.S. GC observations \cite{HESS:2016mib,HESS:2022ygk}. These constraints collectively exclude the red shaded area with $\langle \sigma v\rangle\gtrsim10^{-28}~\cm^3/\rm s$. The red dashed line represents the sensitivities of the future MeV telescopes AMEGO \cite{AMEGO:2019gny,Kierans:2020otl,Caputo:2022xpx},
E-ASTROGAM \cite{e-ASTROGAM:2016bph,e-ASTROGAM:2017pxr} and MAST \cite{Dzhatdoev:2019kay} in probing weak-scale DM, which is derived from \cite{Cirelli:2025qxx}. The future limit is roughly two orders of magnitude lower than the current one. The maximum detection capability is observed at the GeV scale with $\langle \sigma v\rangle\sim 10^{-31}~\cm^3/\rm s$.

In panel (a) of Figure~\ref{FIG:fig3}, as we discussed in Subsection \ref{W-dd}, the three benchmark lines that satisfy the dark matter relic density have couplings $g_\chi$ of order $\mathcal{O}(1)$ at their respective maximum $m_\chi$, leading to a significant cutoff of  $\langle \sigma v \rangle$ since it is independent of $g^\prime$ at this point. Based on this principle,  the $\Delta N_{\mathrm{eff}}$ constraints and the benchmark lines overlap at large $m_\chi$, despite they have different $g^\prime$. After jointly considering the constraints from indirect detection and $\Delta N_{\mathrm{eff}}$,  a conclusion similar to that in Subsection \ref{W-dd}  is drawn: the $Q_\chi = 0.1$ that can be captured by the indirect detection constraint is eventually excluded by the $\Delta N_{\mathrm{eff}}$ constraints, and larger $Q_\chi$ require $m_{\chi} \gtrsim 400$ GeV. 

\subsubsection{Secluded case}

Unlike the resonance scenario,  the present $\langle \sigma v\rangle$ in the secluded scenario is obtained via multiplying Equation~\eqref{Eqn:cz} by the branching ratio into SM. We still choose the electron final state to present the indirect detection constraints.

For the secluded scenario in panel (b) of Figure~\ref{FIG:fig3},  the required annihilation cross section $\langle \sigma v\rangle$ of $\chi\chi\to Z^\prime Z^\prime$  is about $2\times10^{-26}~\cm^3/\rm s$ as the traditional WIMP DM. Under the condition of satisfying the observed relic density, different $Q_\chi$ give  the same $\langle \sigma v \rangle$, and $m_\chi\lesssim20$ GeV  is excluded by current indirect detection constraints. Future experiments favor $m_\chi$ below TeV, but these regions are excluded by the $\Delta N_{\mathrm{eff}}$ constraints, leaving only the region above TeV to survive.

Here we only consider the electron flavor final state.  The other two flavors have almost the same $\langle \sigma v \rangle$, and only the corresponding indirect detection constraints change slightly, but this does not affect the final conclusion. Namely, the constraints from  $\Delta N_{\rm eff}$ on $m_\chi$ are stronger than those from indirect detection.

\subsection{Comprehensive discussion}\label{W-neff}

In this subsection, we discuss, separately for the resonance and secluded scenarios, the parameter space that is allowed by the combined phenomenological constraints and promising to be probed by the future experiments.

\subsubsection{Resonance case}

\begin{figure}
	\begin{center}
		\includegraphics[width=0.45\linewidth]{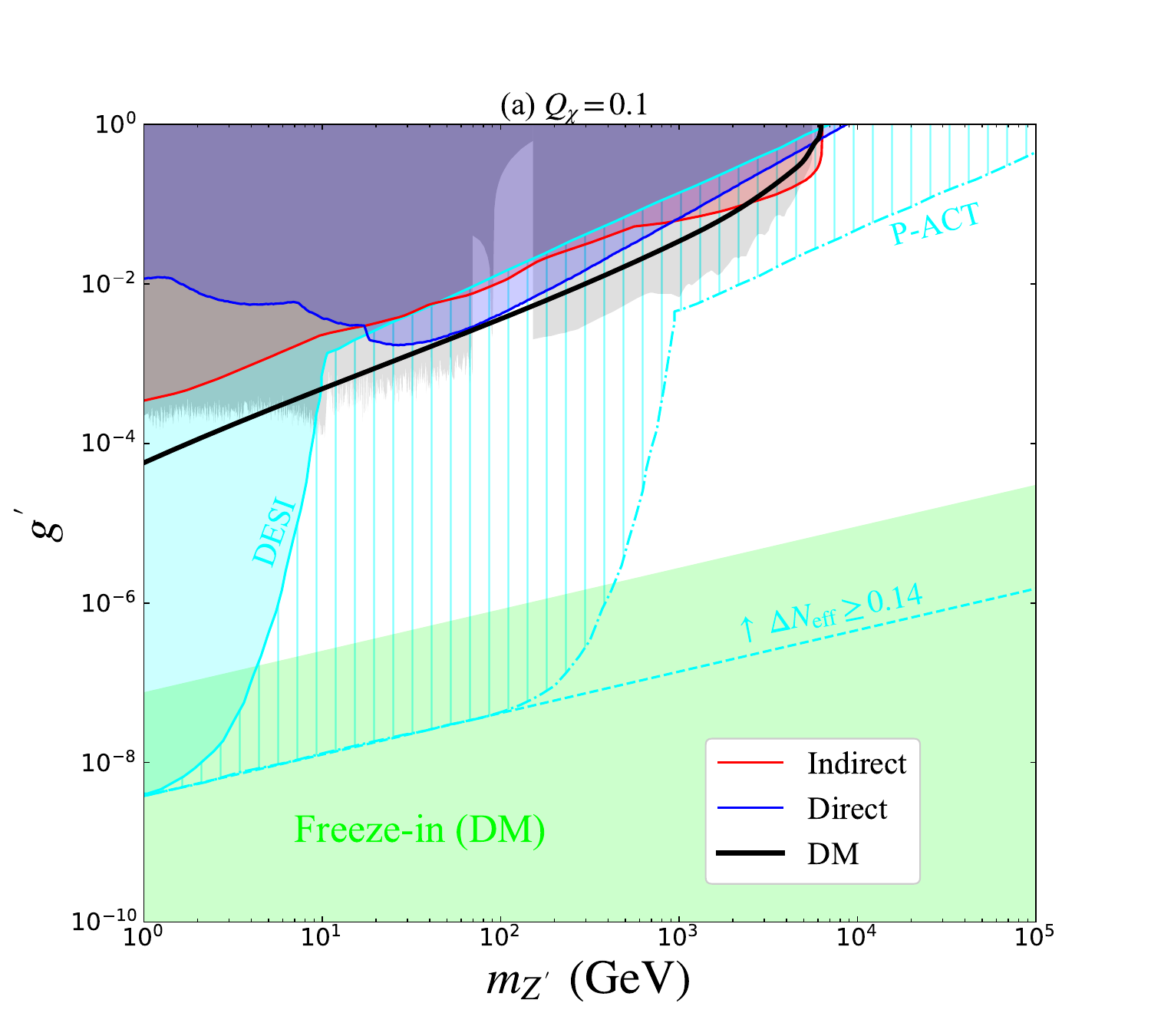}
		\includegraphics[width=0.45\linewidth]{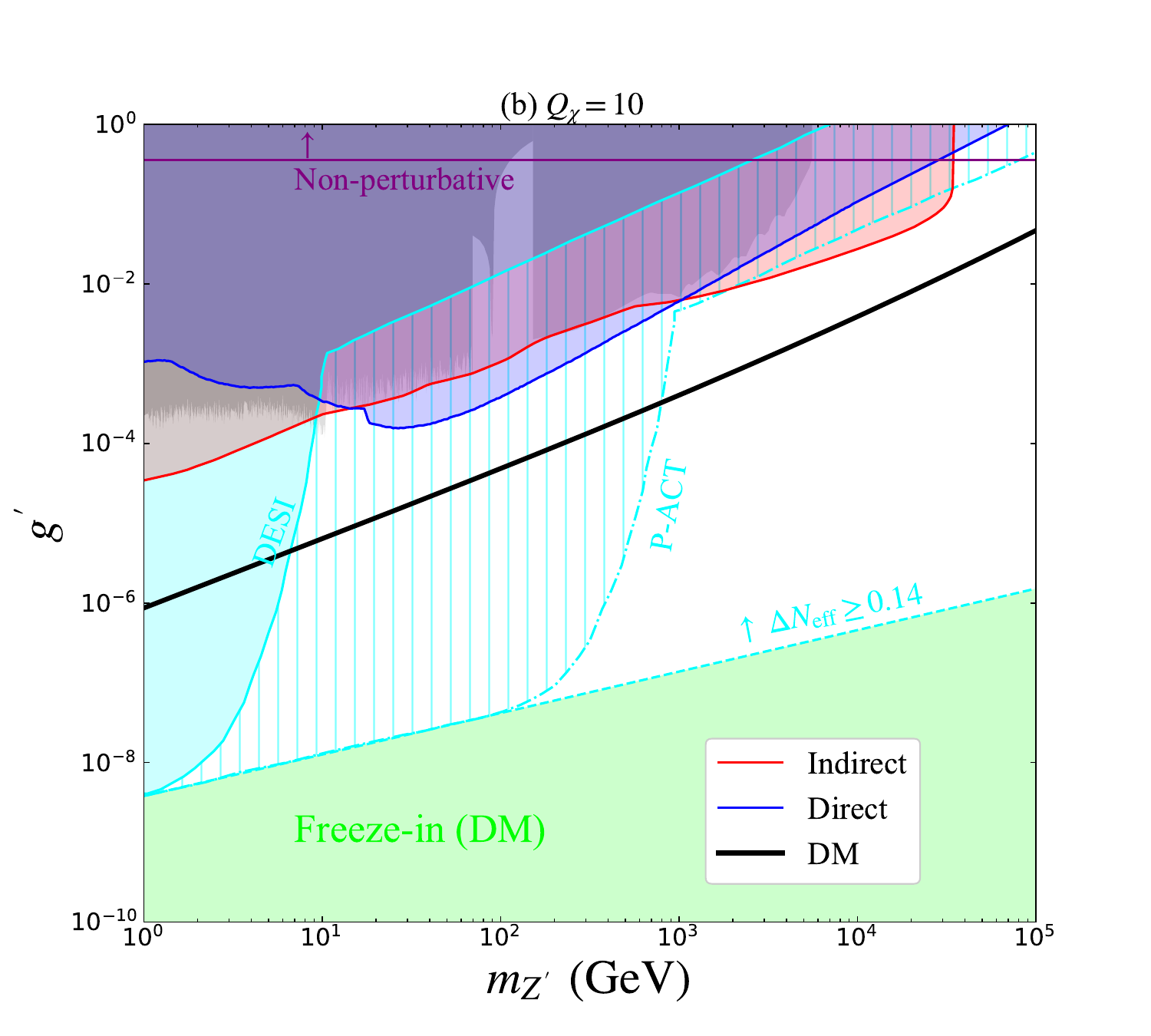}
		\includegraphics[width=0.45\linewidth]{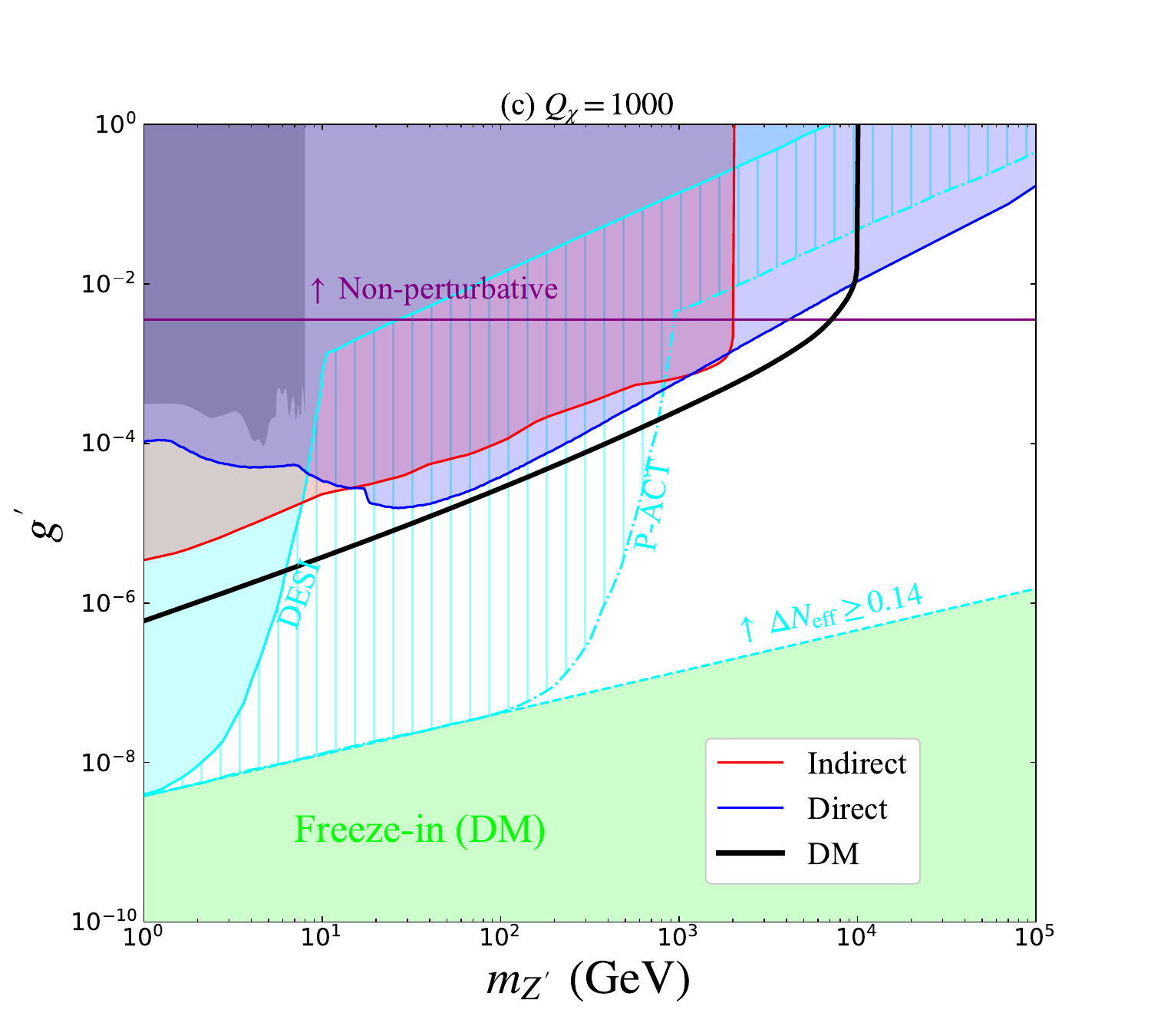}
		\includegraphics[width=0.45\linewidth]{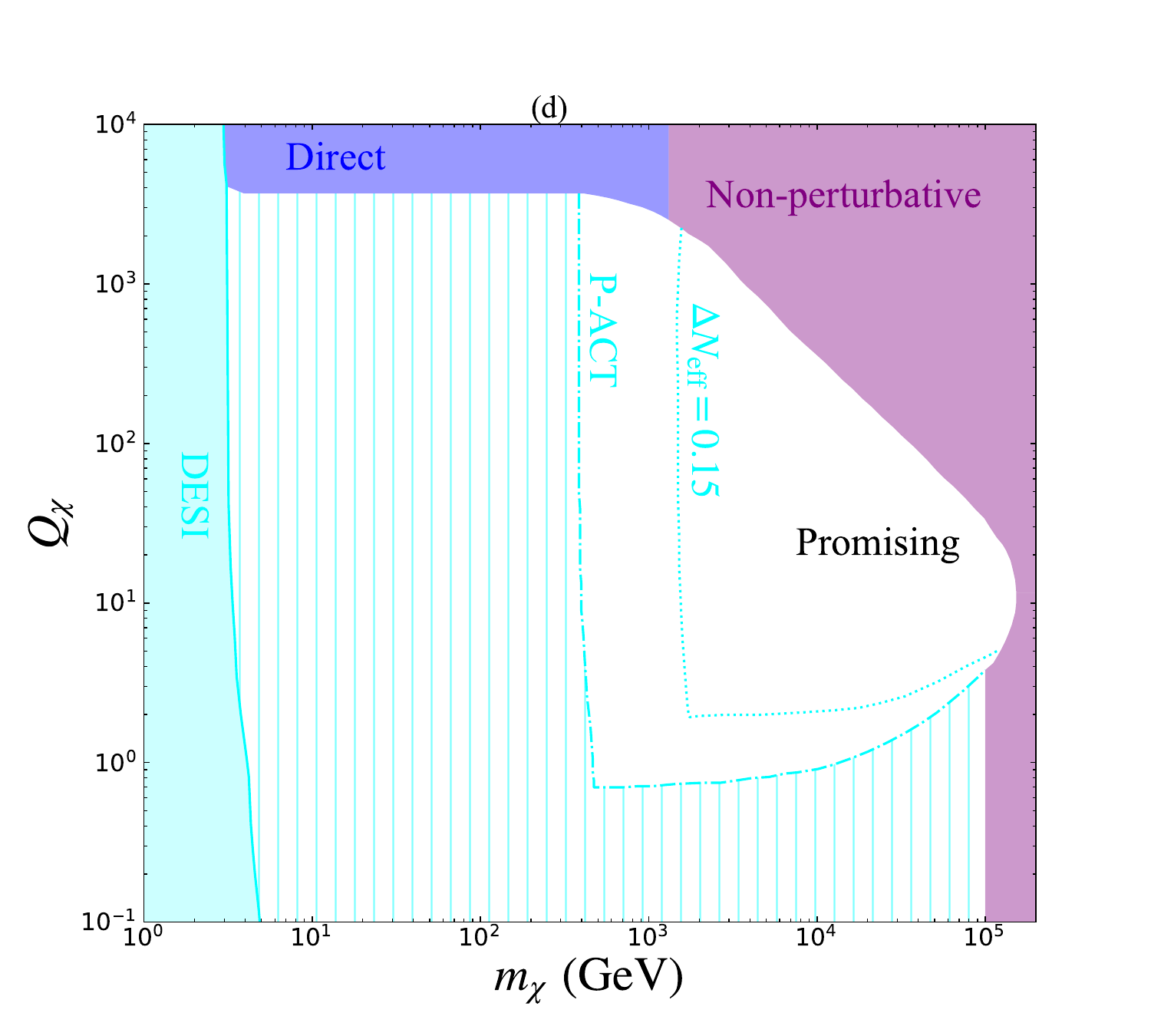}
	\end{center}
	\caption{Comprehensive constraints in the resonance scenario. Panel (a), (b), and (c) correspond to the comprehensive constraints in the $m_{Z^\prime}-g^\prime$ parameter space with $Q_\chi=0.1$, $Q_\chi=10$, and $Q_\chi=1000$, respectively.  Among them, the cyan solid, dot-dashed, and dashed lines represent the $\Delta N_{\mathrm{eff}}$  constraints from  DESI, P-ACT, and thermalization, respectively. The blue and red regions are excluded by DM direct and indirect detection constraints. The parameter space above the purple solid line is non-perturbative of $g_\chi$. The gray region with $g^\prime\gtrsim10^{-4}$ represents the collider constraints induced by the $Z^\prime$. DM is produced non-thermally in the green region with  tiny $g^\prime$. The black lines satisfy the observed relic density of DM with $r_{Z^\prime}=2.001$. Panel (d) shows the  promising parameter space  that can be captured by future  $\Delta N_{\mathrm{eff}}$-related experiments CMB-S4 \cite{Abazajian:2019eic} and CMB-HD \cite{CMB-HD:2022bsz} under comprehensive  constraints. The corresponding colors are consistent with those in panels (a), (b) and (c).
	}
	\label{FIG:fig4}
\end{figure}

\textbf{$\Delta N_{\mathrm{eff}}$  constraints:} In the resonance scenario, when $\nu_R$  decouples from the thermal bath via the freeze-out mechanism.  The decoupling temperature $T_{\rm dec}^{\nu_R}$ is affected by the processes $\nu_R\bar{\nu}_R\to f\bar{f}$ and $\nu_R\bar{\nu}_R\to \chi\bar{\chi}$. However, the latter one is effective only under the condition $T_{\rm dec}^{\chi}<T_{\rm dec}^{\nu_R}$. For WIMP-type dark matter, we take $T_{\rm dec}^{\chi}\simeq m_\chi/25$. According to Equation~\eqref{Eqn:neff-w}, the decoupling temperatures  of $\nu_R$ corresponding to DESI bound $\Delta N_{\eff}\lesssim0.4$ and P-ACT constraint $\Delta N_{\eff}\lesssim0.17$ satisfy $T_{\rm dec}^{\nu_R}\gtrsim0.29$ GeV and $T_{\rm dec}^{\nu_R}\gtrsim40$ GeV, respectively. Based on the critical values,   we obtain the ranges where $\nu_R\bar{\nu}_R\to \chi\bar{\chi}$ takes effect under the DESI  and the  P-ACT constraints are $m_\chi\lesssim7.25$ GeV and $m_\chi\lesssim1000$ GeV, accordingly.  For the validity range of this process, a more detailed discussion is given in the Appendix~\ref{SEC:AEDT}.  The resulting DESI and P-ACT constraints in the $m_{Z^\prime}-g^\prime$ parameter space are shown in Figure~\ref{FIG:fig4} , corresponding to the cyan solid line and dot-dashed line, respectively.

\textbf{Thermalization constraints:} Regarding the thermalization condition of $\nu_R$, as shown in panel (b) of Figure~\ref{FIG:fig1} for the resonance scenario, the maximum reaction rates of $\nu_R\bar{\nu}_R\to f\bar{f}$ and $\nu_R\bar{\nu}_R\to \chi\bar{\chi}$  appear approximately at $T\sim m_{Z^\prime}/3$, which is consistent with that in Ref.~\cite{Abazajian:2019oqj}.  We obtain the thermalization constraint when the maximum reaction rate falls to 1. In the resonance scenario, the decoupling temperature of DM is lower than this value, so the contributions of both processes are taken into account.  The cyan dashed line is  utilized to represent the  thermalization constraints in Figure~\ref{FIG:fig4}. The upper region corresponds to $\Delta N_{\rm eff}\gtrsim0.14$, which arises from the minimum value derived in Equation~\eqref{Eqn:neff-w} with three generations of $\nu_R$.  Furthermore, we use the annihilation cross section in Equation~\eqref{Eqn:cf-on} to calculate the thermalization constraints of DM, below which, the green region with tiny $g^\prime$ in Figure~\ref{FIG:fig4} indicates that DM cannot reach thermal equilibrium, and it is produced via the freeze-in mechanism. We will discuss the freeze-in scenario in detail in Section \ref{SEC:FIMP}.

Under the cooperation of thermalization limits, the parameter space of $g^\prime\gtrsim\mathcal{O}(10^{-8})$ with GeV scale $m_{Z^\prime}$  is not allowed by the DESI constraint, while the exclusion capability of P-ACT on $m_{Z^\prime}$  is about two orders of magnitude stronger than that of DESI. We report that the thermally produced $\Delta N_{\rm eff}$ constraints remain nearly unchanged across different $Q_\chi$ cases in Figure~\ref{FIG:fig4}. The primary reason for not changing is that,  as shown in Figure~\ref{FIG:fig1} (b),  the reaction rate of $\nu_R\bar{\nu}_R\to \chi\bar{\chi}$ as the dominant process is almost identical to that of $\nu_R\bar{\nu}_R\to f\bar{f}$ as the dominant process.  Furthermore, when $\chi$ decouples before $\nu_R$ at large mass, the process $\nu_R\bar{\nu}_R\to \chi\bar{\chi}$ does not contribute to $\Delta N_{\rm eff}$.  Taking the P-ACT constraint in the resonance scenario with $Q_\chi=1000$ as an example, the reason it remaining continuously varying at $m_\chi\simeq1000$ GeV is that the constraint is determined by the maximum $x_{Z'}$ segment of the piecewise Equation~\eqref{Eqn:vrf} after the inflection point $(m_{Z^\prime}=985~\GeV, g^\prime=4.8\times10^{-3})$,  and this equation is unaffected by $Q_\chi$.

\textbf{DM direct and indirect constraints:} In addition, as shown by the blue and red regions in Figure~\ref{FIG:fig4}, we reproduce the  constraints of current DM direct and indirect detection experiments  in the $m_{Z^\prime}-g^\prime$ parameter space by using  Equation~\eqref{Eqn:dd} and Equation~\eqref{Eqn:r-id}, respectively. The lower bound of the excluded region decreases as $Q_\chi$ increases. 

\textbf{Perturbative constraint:} For events with  large $Q_{\chi}$,  $g_\chi$ becomes non-perturbative if it has a relatively large magnitude, which corresponds to the purple region in the figures, namely, $g_{\chi} > \sqrt{4\pi}$.  

\textbf{Collider constraints:} In the aspect of collider constraints,  depending on the decay mode  of $Z^\prime$, colliders could search for  $Z^\prime$  through the visible or the invisible final states, and their corresponding limits are distinctly different. In our work, the visible  decay of $Z^\prime$ is dominant in the resonance scenario with $Q_\chi=0.1$  and 10. Hence we adopt the relevant  constraints from the current experiments BaBar \cite{BaBar:2014zli}, LHCb \cite{LHCb:2017trq,LHCb:2019vmc}, LEP \cite{KA:2023dyz,ALEPH:2013dgf},  as well as CMS and ATLAS  \cite{CMS:2021ctt,ATLAS:2019erb}. These experimental limits collectively exclude  the shaded gray area with $g^\prime \gtrsim10^{-4}$ in Figure~\ref{FIG:fig4} and Figure~\ref{FIG:fig5}. The invisible decay  $Z^\prime\to\chi\bar{\chi}$ is dominant in the resonance scenario with $Q_\chi=1000$. BaBar searches for the invisible decays of $Z^\prime$ \cite{BaBar:2017tiz}. The results are shown as the shaded region in panel (c) of Figure~\ref{FIG:fig4}, which  excludes the parameter space with $m_{Z^\prime}\lesssim10$ GeV and $g^\prime \gtrsim10^{-4}$.  For the  constraints of future colliders, such as Belle II \cite{Ferber:2022ewf,Dolan:2017osp} and FCC-ee  \cite{Karliner:2015tga}, although a wider search range could be tested, almost all sensitivities lie within the current P-ACT limit  just like the current constraints. So we do not show the corresponding future collider sensitivities.

\textbf{Comprehensive discussion:} In the resonance scenario, based on the comprehensive  constraints on specific benchmark points that satisfy the DM relic density in panel (a), (b) and (c) of Figure~\ref{FIG:fig4},  we obtain the promising parameter space in the $m_\chi-Q_\chi$ plane that  evades all current constraints in panel (d) of Figure~\ref{FIG:fig4}. The dark matter and $Z^\prime$ within it can be captured by at least one type of future experiments. Specifically, the P-ACT constraint determines the lower limit of the promising region, which roughly requires $Q_\chi\gtrsim0.7$ and $m_\chi\gtrsim 400$ GeV. The upper limit of $m_\chi$ is determined by the perturbative constraint, reaching a maximum value $1.54\times10^5$ GeV  when $Q_\chi\simeq10$.  The dark matter direct detection  limits the maximum value of $Q_\chi$ up to 3700. And the hopeful $g^\prime$ decreases from $\mathcal{O}(0.1)$ to $\mathcal{O}(10^{-4})$ as $Q_\chi$ increases. For the looser DESI constraint, the lower limit of $m_\chi$ could be as small as 3 GeV.   In general, a smaller $\Delta N_{\rm eff}$ requires a larger $m_\chi$ and $Q_\chi$. For example, $\Delta N_{\rm eff}\lesssim0.15$ corresponds to $m_\chi\gtrsim1500$  GeV and $Q_\chi\gtrsim2$. It should be noted that the minimum  achievable $\Delta N_{\rm eff}$ in this scenario is 0.14.  Future CMB-S4 ($\Delta N_{\rm eff}\lesssim0.06$)  and CMB-HD ($\Delta N_{\rm eff}\lesssim0.027$) will  cover this promising region. Meanwhile, according to the results in Figure~\ref{FIG:fig2} and Figure~\ref{FIG:fig3}, the TeV scale $m_\chi$ could also be captured by both future direct and indirect detection experiments.

\subsubsection{Secluded case}
   
\begin{figure}
	\begin{center}
		\includegraphics[width=0.45\linewidth]{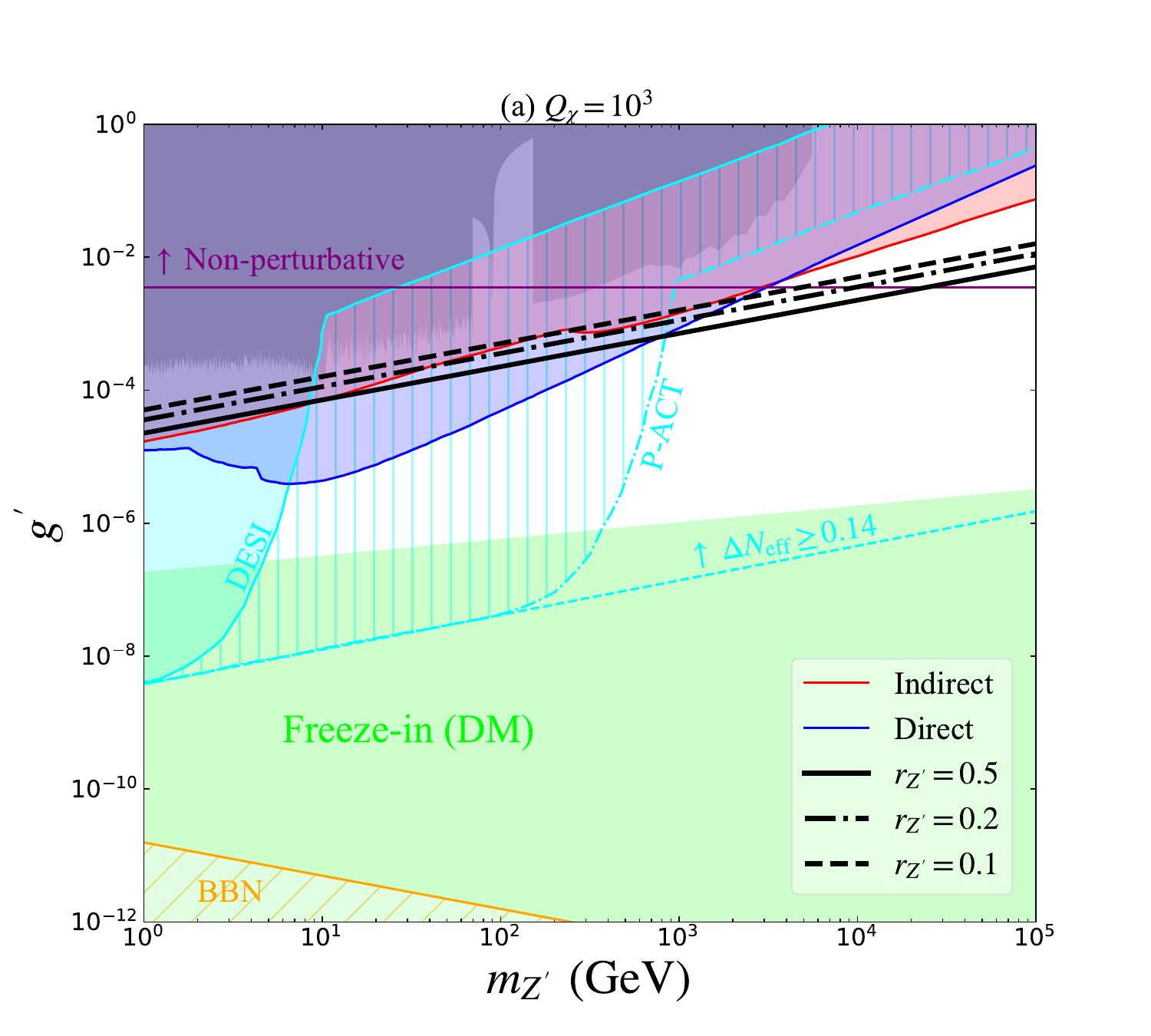}
		\includegraphics[width=0.45\linewidth]{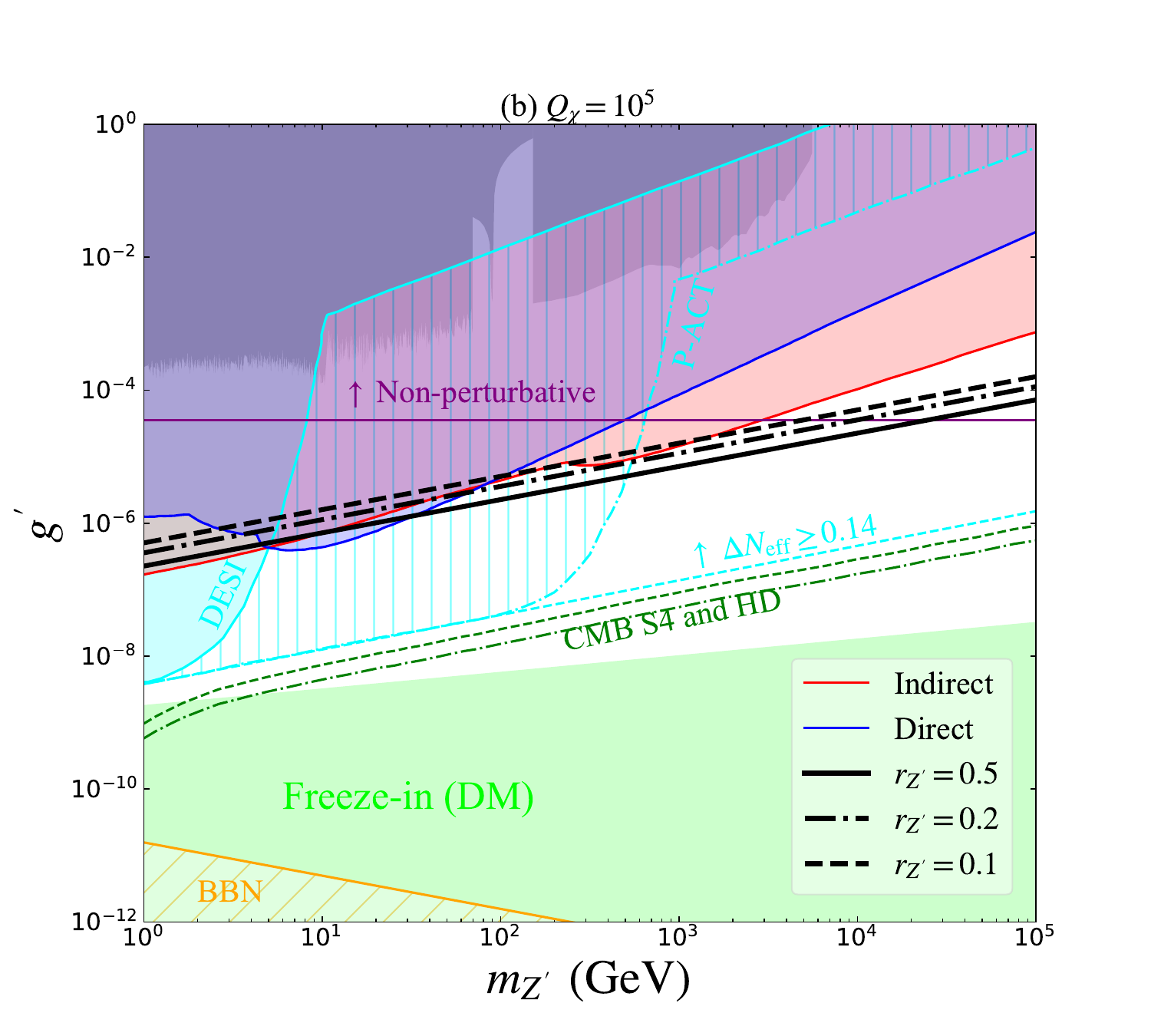}
		\includegraphics[width=0.45\linewidth]{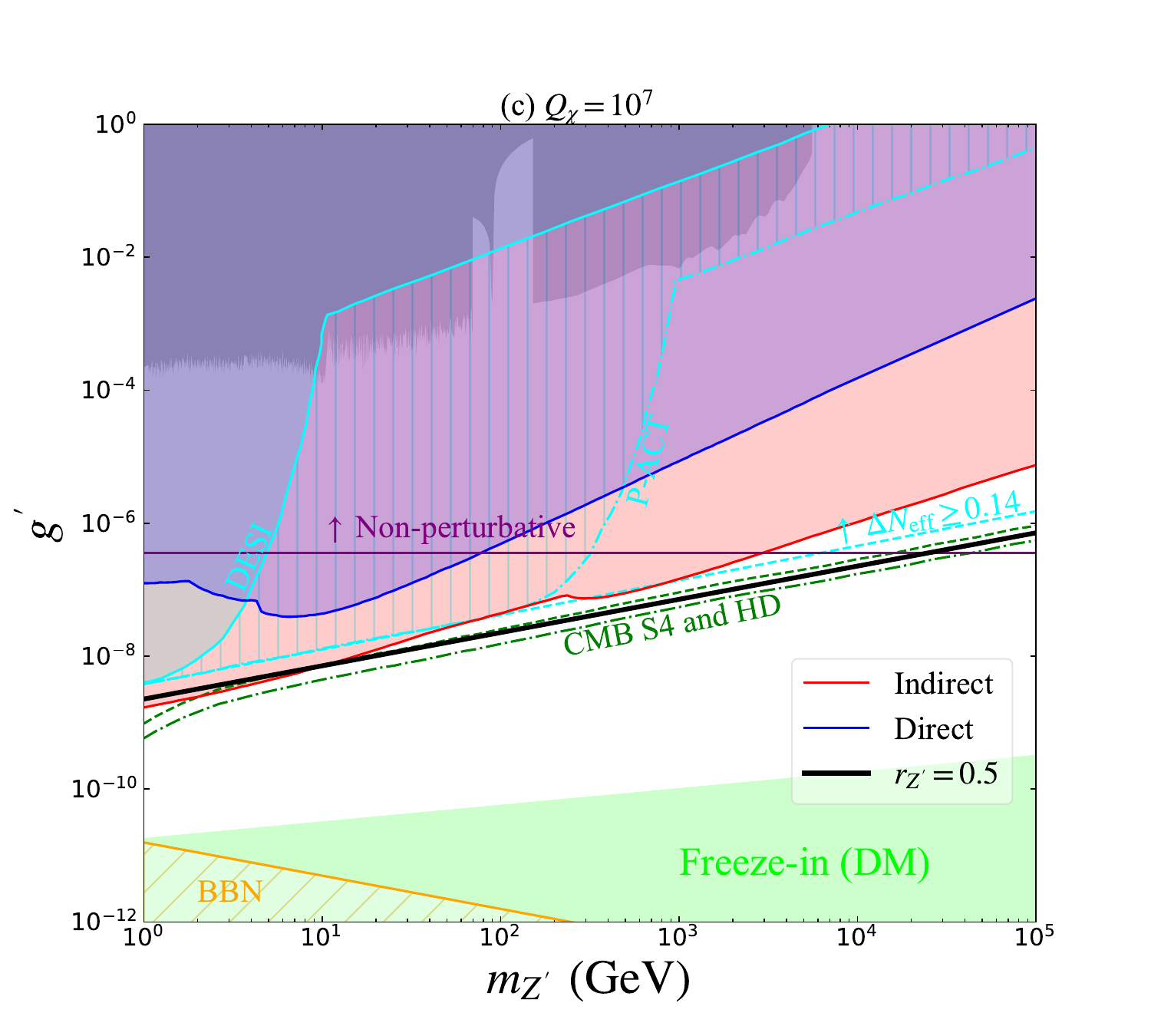}
		\includegraphics[width=0.45\linewidth]{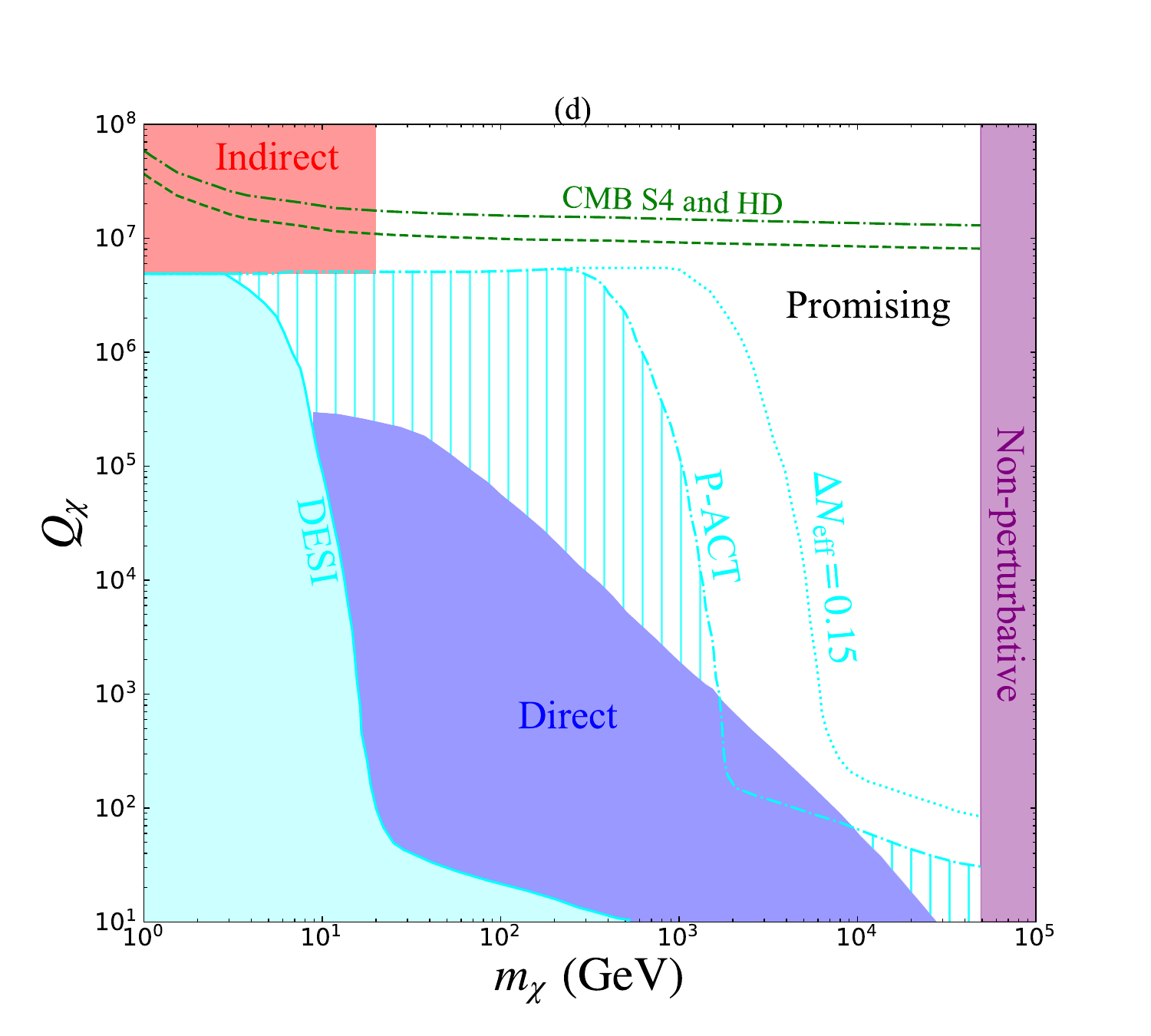}
	\end{center}
	\caption{Same as Figure~\ref{FIG:fig4} but for the secluded scenario. Panels (a), (b), and (c) correspond to cases $Q_\chi=10^3$, $Q_\chi=10^5$, and $Q_\chi=10^7$, respectively. In each case, the black solid, dot-dashed, and dashed lines represent the benchmark $r_{Z^\prime}=0.5$, 0.2, and 0.1, respectively.  For clarity, in panel (c) we only show the benchmark with $r_{Z^\prime}=0.5$, as the results of $r_{Z^\prime}=0.2$ and 0.1 largely overlap  with $\Delta N_{\rm eff}=0.14$. The orange shaded region indicates that the lifetime of $Z^\prime$ is longer than 0.1 s, which would be constrained by BBN observations. In panels (b) and (c), the thermalization range of dark matter will reach the non-thermal domain of $\nu_R$, where $\nu_R$ is generated through the freeze-in mechanism \cite{Adshead:2022ovo}. The green dashed and dot-dashed curves correspond to the CMB-S4 and CMB-HD constraints. The promising region in panel (d) is obtained with $r_{Z^\prime}=0.5$.
	}
	\label{FIG:fig5}
\end{figure}

\textbf{$\Delta N_{\mathrm{eff}}$  and  Thermalization constraints:} 
In the secluded scenario,  due to the suppression of $\nu_R\bar{\nu}_R\to \chi\bar{\chi}$ with off-shell $Z^\prime$, $\nu_R\bar{\nu}_R\to f\bar{f}$ is completely dominant for the $\Delta N_{\mathrm{eff}}$  constraints and thermalization constraints of $\nu_R$.  Adopting the same approach as in the resonance scenario, we show the  DESI and P-ACT results in Figure~\ref{FIG:fig5} by the cyan solid line and dot-dashed line, respectively. While the cyan dashed line denotes the thermalization constraint of $\nu_R$. Both of them remain almost unchanged compared with those in the resonance scenario.
Whereas the thermalization constraints of dark matter are obtained from Equation~\eqref{Eqn:cz}, which is lower than the thermalization constraints of $\nu_R$ when $Q_\chi \geq10^4$, implying that  $\nu_R$ should be produced non-thermally in  the middle region. Since DM contributes negligibly to $\nu_R$ in this scenario, $\nu_R$ is produced non-thermally through $f\bar{f}\to\nu_R\bar{\nu}_R$ \cite{Adshead:2022ovo}. The corresponding CMB-S4 and CMB-HD constraints are shown as green dashed and dot-dashed curves in panel (b) and (c) of Figure~\ref{FIG:fig5}.

\textbf{DM direct and indirect constraints:} In this scenario, the current DM direct and indirect detection constraints in $m_{Z^\prime}-g^\prime$ parameter space are  reproduced through the Equation~\eqref{Eqn:dd} and Equation~\eqref{Eqn:cz}, respectively. They exclude the blue and red regions in Figure~\ref{FIG:fig5}, respectively. 

\textbf{Perturbative and Collider constraints:}  The purple region in Figure~\ref{FIG:fig5} represents the perturbative constraints $g_{\chi} > \sqrt{4\pi}$.  In this scenario, $Z^\prime$ decays predominantly into visible final states, therefore, the collider constraints applied are identical to those from the cases of $Q_\chi=0.1$  and 10  in the resonance scenario, which are shown as the gray shaded region in Figure~\ref{FIG:fig5}.

\textbf{Comprehensive discussion:} In the secluded scenario, the promising region is constrained to be smaller  as $r_{Z^\prime}$ decreases. Therefore, we fix $r_{Z^\prime}=0.5$ to obtain a relatively larger space in panel (d) of Figure~\ref{FIG:fig5}.  Thermally produced $\Delta N_{\rm eff}$ could only occur at $Q_\chi\lesssim5\times10^6$.  Under the P-ACT constraint, the promising region satisfies $m_\chi\gtrsim 250$~GeV and $Q_\chi\gtrsim30$.  The minimum $m_\chi$ appears at the intersection  of thermalization and the P-ACT constraint, and the perturbative  bound determines the upper limit of $m_\chi$ as $4.9\times10^4$ GeV.  Different $\Delta N_{\rm eff}$ determines different lower limits of  $m_\chi$ and $Q_\chi$. For example, $m_\chi\gtrsim 2.7$ GeV can satisfy the looser DESI constraint, but the direct detection constraints exclude the region with $Q_\chi\lesssim3\times10^4$, which almost lies entirely within the P-ACT limit.   A smaller $\Delta N_{\rm eff}=0.15$ corresponds to the minimum  $m_\chi$ and $Q_\chi$ being 1000 GeV and 90, respectively.

When $Q_\chi\gtrsim5\times10^6$, the corresponding $g'$ for correct relic density is too small, so $\nu_R$ could only be produced via the freeze-in mechanism. The matching  CMB-S4 and CMB-HD constraints are located at $Q_\chi\sim\mathcal{O}(10^7)$, which corresponds to the minimum  detectable $g^\prime\sim\mathcal{O}(10^{-9})$. They both are promising to probe $m_\chi\gtrsim20$ GeV under the constraints of indirect detection. For the  parameter space above the CMB-HD constraints, there is a tiny $\Delta N_{\rm eff}$ that is difficult to  be captured by future experiments. On the other hand, the corresponding $g^\prime$ might be too small. For instance, $g^\prime\lesssim\mathcal{O}(10^{-11})$ is disallowed by the BBN constraint, which requires the lifetime of $Z^\prime$ to be less than 0.1 s.  Furthermore, when $Q_\chi\lesssim10^4$, $m_\chi$ above TeV can be doubly checked by future direct detection experiments.

In comparison, the secluded scenario is more promising for detection than the resonance one. Once future experiments detect $\Delta N_{\rm eff}$ below 0.14, the resonance scenario will be disfavored. Moreover, the secluded scenario could even produce extremely small $\Delta N_{\rm eff}$ that goes beyond the sensitivity of CMB-HD, which makes it consistently promising for longer-term experiments.

\section{FIMP scenario}\label{SEC:FIMP}

\subsection{Relic density and $\Delta N_{\eff}$}\label{F-rd}

When $\chi$ and $\nu_R$ can not reach thermal equilibrium, their abundances $Y_\chi$ and $Y_{\nu_R}$ are generated through the freeze-in mechanism. The corresponding Boltzmann equations are
\begin{eqnarray}\label{Eqn:y-f}
	\frac{dY_\chi}{dx_{Z^\prime}} &=& \frac{s}{\mathcal{H}x_{Z^\prime}} \langle \sigma v\rangle_{f\bar{f}\to\chi\bar{\chi}}\left((Y_{f}^\eq)^2-\frac{(Y_{f}^\eq)^2}{(Y_{\chi}^{\eq})^2}Y_{\chi}^2\right),\\
	\frac{dY_{\nu_R}}{dx_{Z^\prime}} &=& \frac{s^{2/3}}{\mathcal{H}x_{Z^\prime}} m_{Z^\prime}\langle \sigma v\rangle_{f\bar{f}\to\nu_R\bar{\nu_R}}(Y_{\nu_R}^\eq)^2,
\end{eqnarray}
where we neglect the  contributions of the $t$-channel $Z^\prime Z^\prime\to\chi\bar{\chi},\nu_R\bar{\nu_R}$, since they are strongly suppressed by the fourth power of the couplings.  The conversion process $\chi\bar{\chi}\to\nu_R\bar{\nu_R}$ is also neglected, which has a very small reaction rate and does not affect the final results at all under our verification. In addition, as shown in Figure~\ref{FIG:fig8} (b),  ignoring the derivative of the SM dofs in these Boltzmann equations has tiny impact on  our conclusions. The notations for the parameters could be found in the WIMP scenario. We refer to the calculation rule for $\Delta N_{\rm eff}$ in Ref.~\cite{Biswas:2022vkq}, namely
\begin{eqnarray}\label{Eqn:neff-f}
	\Delta N_{\rm eff}=2\times3\times\left(\frac{\rho_{\nu_R}}{\rho_{\nu_L}}\right)_{\rm CMB}=6 \times\left(\frac{s^{4/3}Y_{\nu_R}}{\rho_{\nu_L}}\right)_{T=10~\MeV}
\end{eqnarray}
where $\rho_{\nu_L}=7\pi^2T^4/120$.

\begin{figure}
	\begin{center}
		\includegraphics[width=0.45\linewidth]{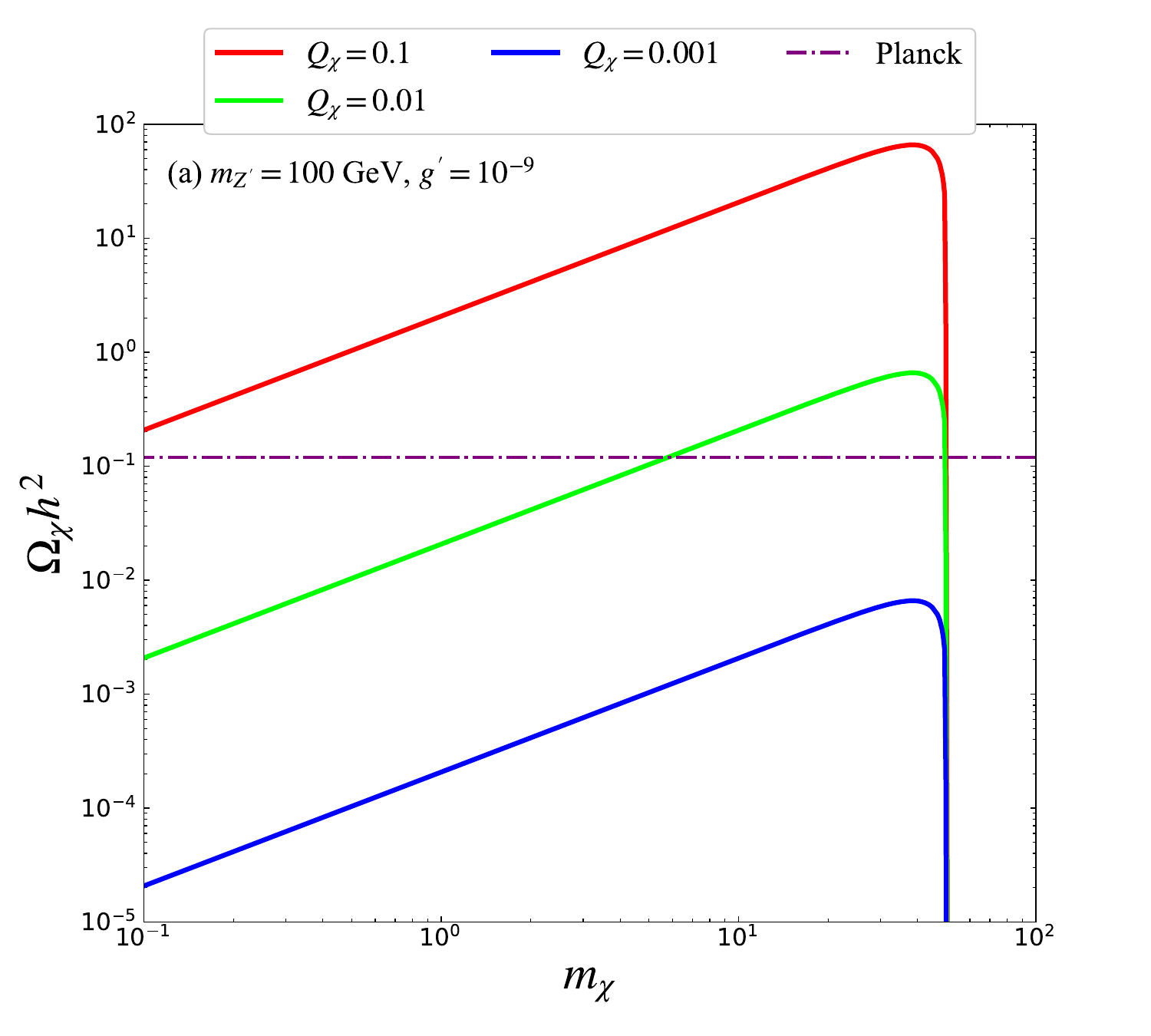}
		\includegraphics[width=0.45\linewidth]{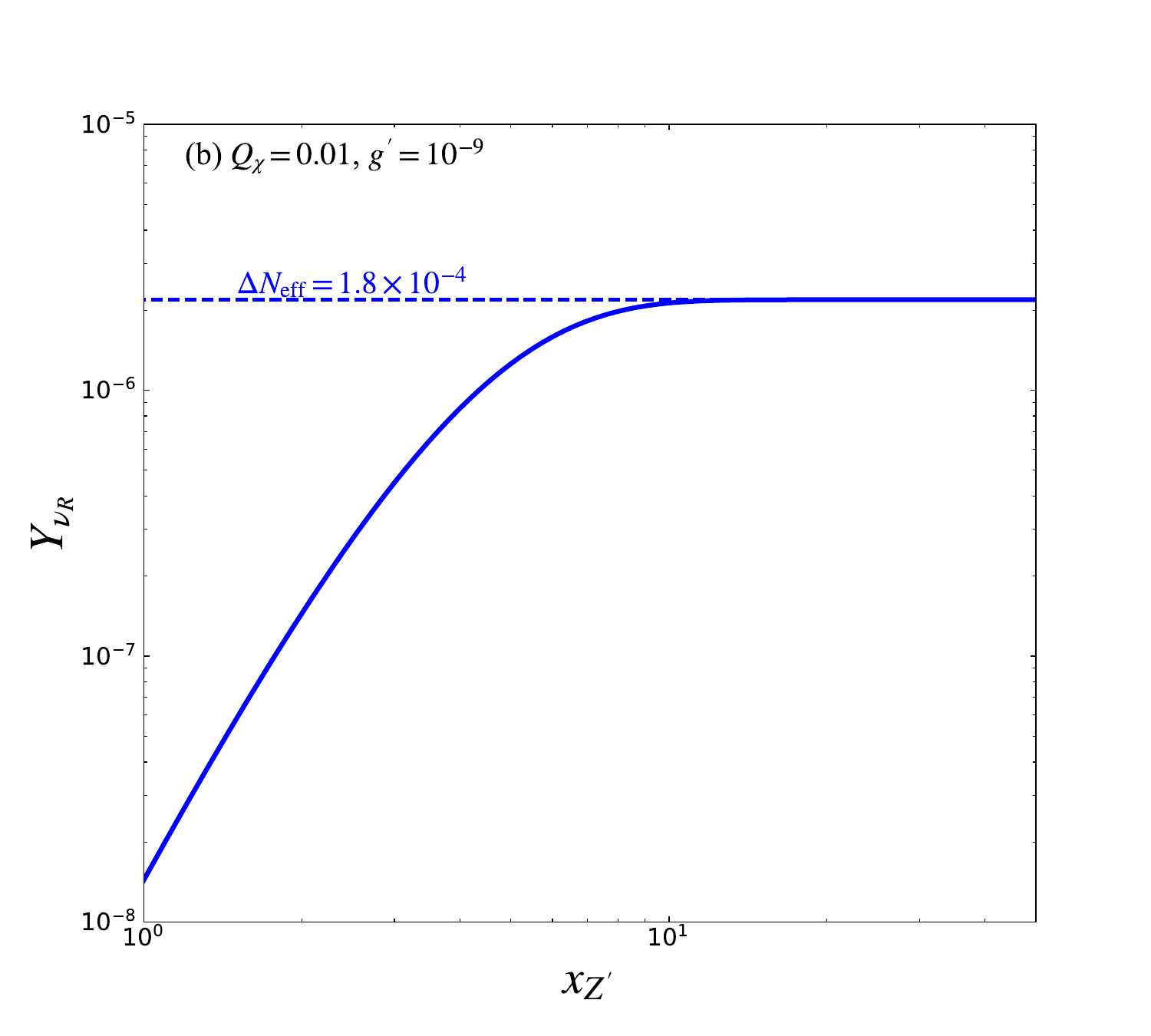}
		\includegraphics[width=0.45\linewidth]{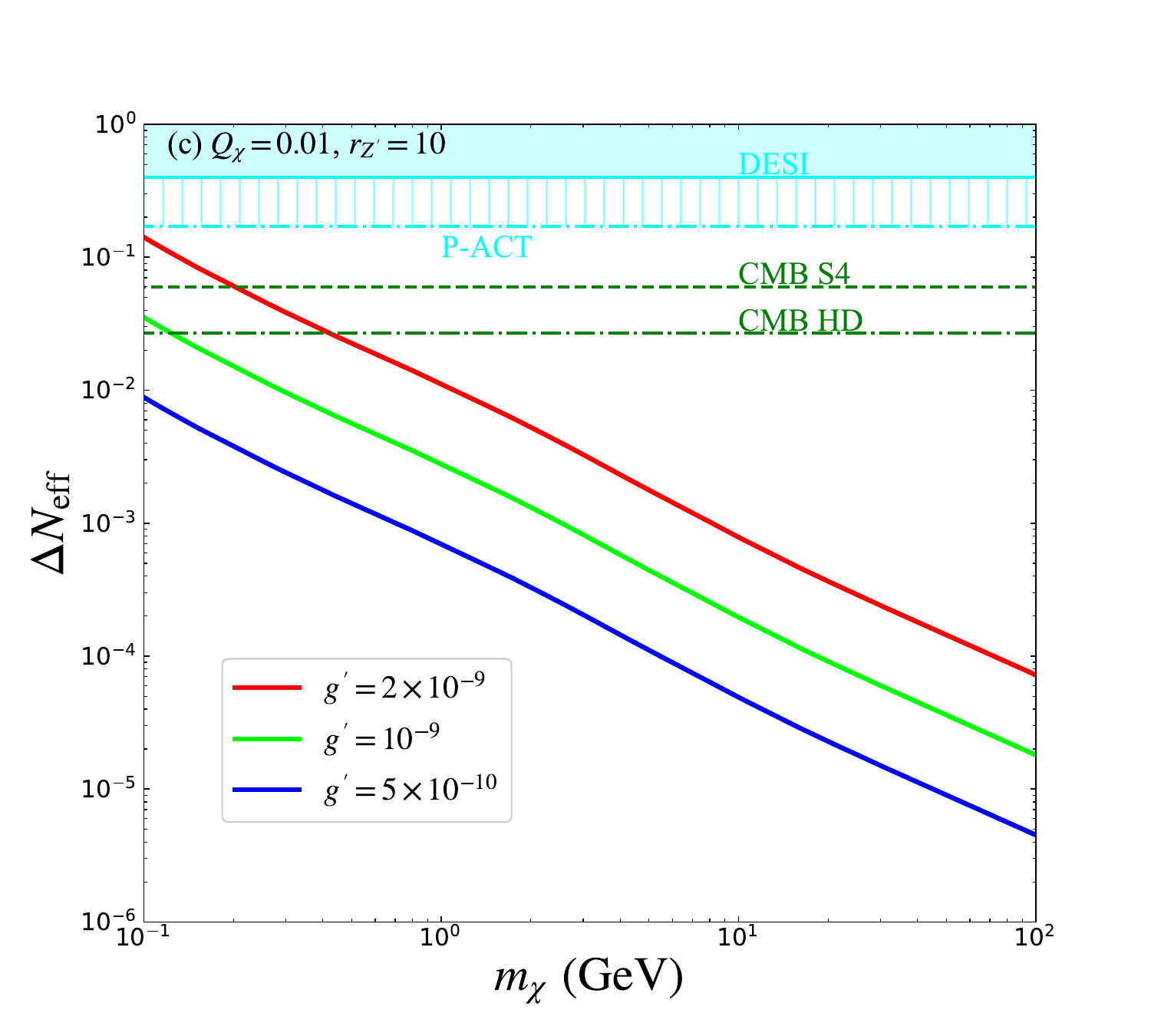}
	\end{center}
	\caption{ Panel (a): The dependence of DM relic density in the FIMP scenario. Panel (b): The evolution of $Y_{\nu_R}$ with a correct DM relic density. Panel (c): The dependency of $\Delta N_{\eff}$ on $m_\chi$ in the FIMP scenario.  In panel (a) with fixed $m_{Z^\prime}=100$ GeV and $g^\prime=10^{-9}$, the red, lime, and blue solid lines represent $Q_\chi=0.1$, 0.01, and 0.001, respectively. The purple dot-dashed line represents the Planck observation of dark matter. In panel (b),  $m_\chi\simeq6$~GeV with $Q_\chi=0.01$ yields a DM relic density consistent with observations for $m_{Z^\prime}=100$ GeV and $g^\prime=10^{-9}$ in panel (a). The resulting $Y_{\nu_R}$ corresponds to $\Delta N_{\eff} =1.8\times10^{-4}$. In panel (c), the red, lime, and blue benchmarks have $g^\prime=2\times10^{-9}$, $g^\prime=10^{-9}$, and $g^\prime=5\times10^{-10}$, respectively. The constraints on $\Delta N_{\eff}$ are consistent with those in Figure~\ref{FIG:fig1} (c).
	}
	\label{FIG:fig6}
\end{figure}

Provided $m_{Z'}>2m_\chi$, the DM is dominantly produced via the two-body decay $Z'\to \chi \bar{\chi}$. The corresponding decay rate is in Equation \eqref{Eqn:zpxx}. The out of equilibrium condition  $\mathcal{K}_1(T)~\Gamma_{Z^\prime\to \chi\bar{\chi}}/\mathcal{K}_2(T)\lesssim  \left. \mathcal{H}(T)\right|_{T=m_{Z^\prime}}$ roughly implies 
\begin{equation}\label{Eqn:gx}
	g_\chi=g'\times Q_\chi \lesssim 4\times10^{-8}\times \left(\frac{m_{Z'}}{100~\text{GeV}}\right)^{1/2}.
\end{equation}
which corresponds to the red region in Figure~\ref{FIG:fig7} .

In panel (a) of Figure~\ref{FIG:fig6}, the dark matter relic density as a function of $m_\chi$ is shown with $m_{Z^\prime}=100$~GeV and $g^\prime=10^{-9}$, which is clearly in the non‑thermal DM regime.  In the on‑shell regime of $Z^\prime$, the DM relic density $\Omega_{\chi}h^2$ rises with  increasing $m_\chi$, and it increases by two orders of magnitude with a one order of magnitude decrease in $Q_\chi$. The main reason is that $ \langle \sigma v\rangle_{f\bar{f}\to\chi\bar{\chi}}$ is proportional to $\Gamma_{Z^\prime\to \chi\bar{\chi}}$. Actually, for $m_\chi\ll m_{Z^\prime}$, the  dark matter production is effective for $x_{Z^\prime}\lesssim 1$, namely, $x_\chi\ll 1$. Based on the Equation~\eqref{Eqn:cf-on},  we could  make the following approximation \cite{Okada:2020cue}:
	\begin{eqnarray}\label{Eqn:cf-fimp}
			\langle\sigma v \rangle^{\rm FIMP}_{\chi\bar{\chi} \to f\bar{f}} &\simeq& \frac{\pi }{32} (Q_\chi g^\prime)^2\left(\frac{m_{Z^\prime}^3}{m_\chi^5} \right)\mathcal{K}_1 \left(\frac{ m_{Z^\prime}}{m_\chi} x_\chi \right) x_\chi^5  \nonumber \\
			&\simeq& \left(\frac{\pi}{32}\right) \left(Q_\chi g^\prime\right)^2 \left(\frac{x_{Z^\prime}^4}{m_{Z^\prime}^2}\right)
	\end{eqnarray}
for $x_\chi\lesssim m_{\chi} /m_{Z^\prime} $, while $\langle\sigma v \rangle^{\rm FIMP}_{\chi\bar{\chi} \to f\bar{f}}  \simeq 0$ for $x_\chi > m_\chi/m_{Z^\prime}$ where we neglect the production of dark matter. Here  $\mathcal{K}_1 (y) \propto e^{- y}$ with  $y \geq 1$ and $K_1 (y) \sim 1/y$ with $y \leq 1$ are taken into account.  Within this approximation, the Boltzmann Equation~\eqref{Eqn:y-f} can be solved analytically from $x_{Z^\prime}=x_{rh}$ to  $x_{Z^\prime}=1$ \cite{Okada:2020cue} with the initial condition $Y_\chi(x_{rh})=0$:
\begin{equation}
Y_\chi^{0} \sim Y_\chi(x_{Z^\prime}=1) \simeq 8.5 \times 10^{-6} \left(Q_\chi g^\prime\right)^2 \left(\frac{m_{\rm pl}}{m_Z^\prime}\right).
\end{equation}
Based on the observations of dark matter, the corresponding approximate analytical relation is \cite{Nath:2021uqb,Okada:2020cue}
\begin{equation}\label{Eqn:rd}
	g_\chi=g^\prime\times Q_\chi\simeq2.5\times10^{-12}\left(\frac{m_{Z'}}{m_\chi}\right)^{1/2}
\end{equation}

Once entering the off‑shell regime $m_{Z'}<2m_\chi$, $\Omega_{\chi}h^2$ drops sharply to $\mathcal{O}(10^{-16})$, which is significantly lower than the observational results.  A distinct solution matching the dark matter relic density is $m_\chi\simeq6$ GeV with $Q_\chi=0.01$.   On this basis, we obtain the corresponding $\Delta N_{\eff} =1.8\times10^{-4}$ in panel (b) of Figure~\ref{FIG:fig6}.

In panel (c) of Figure~\ref{FIG:fig6}, we present the relation between $\Delta N_{\eff}$ and $m_\chi$ under various $g^\prime$ for the benchmark choice of  $r_{Z^\prime}=10$ and $Q_\chi=0.01$.  Same as in the WIMP scenario, we do not impose that the benchmarks reproduce the observed relic density. For fixed mass ratio $r_{Z'}=10$, an increasing in $m_{\chi}$ reduces $\langle \sigma v\rangle_{f\bar{f}\to\nu_R\bar{\nu_R}}$, which consequently lowers $\Delta N_{\eff}$. Meanwhile, the resulting $\Delta N_{\eff}$ is proportional to ${g^\prime}^2$.  In the non‑thermal regime where ${g^\prime}\lesssim\mathcal{O}(10^{-9})$, the resulting $\Delta N_{\eff}$ is less than 0.1 when $m_\chi\gtrsim0.1$~GeV, thus avoids exclusion by current P-ACT bounds. The future CMB-HD experiments will be able to test  $m_\chi\lesssim0.4$~GeV when ${g^\prime}=2\times10^{-9}$, whereas ${g^\prime}\lesssim\mathcal{O}(10^{-10})$ is beyond the future reach. By including the quantum-statistical, out-of-equilibrium, and finite lifetime of gauge boson $Z'$ effects, the more precise results can be obtained for the same process $f\bar{f}\to\nu_R\bar{\nu_R}$ \cite{Adshead:2022ovo}, which differs from our results by $\mathcal{O}(10\%)$ on $g^\prime$ with the same  $\Delta N_{\rm eff}$. For the sake of precision, we adopt the $\Delta N_{\rm eff}$ results in Ref. \cite{Adshead:2022ovo} in the subsequent discussion.

\subsection{Comprehensive discussion}\label{F-neff}

\begin{figure}
  	\begin{center}
  		\includegraphics[width=0.45\linewidth]{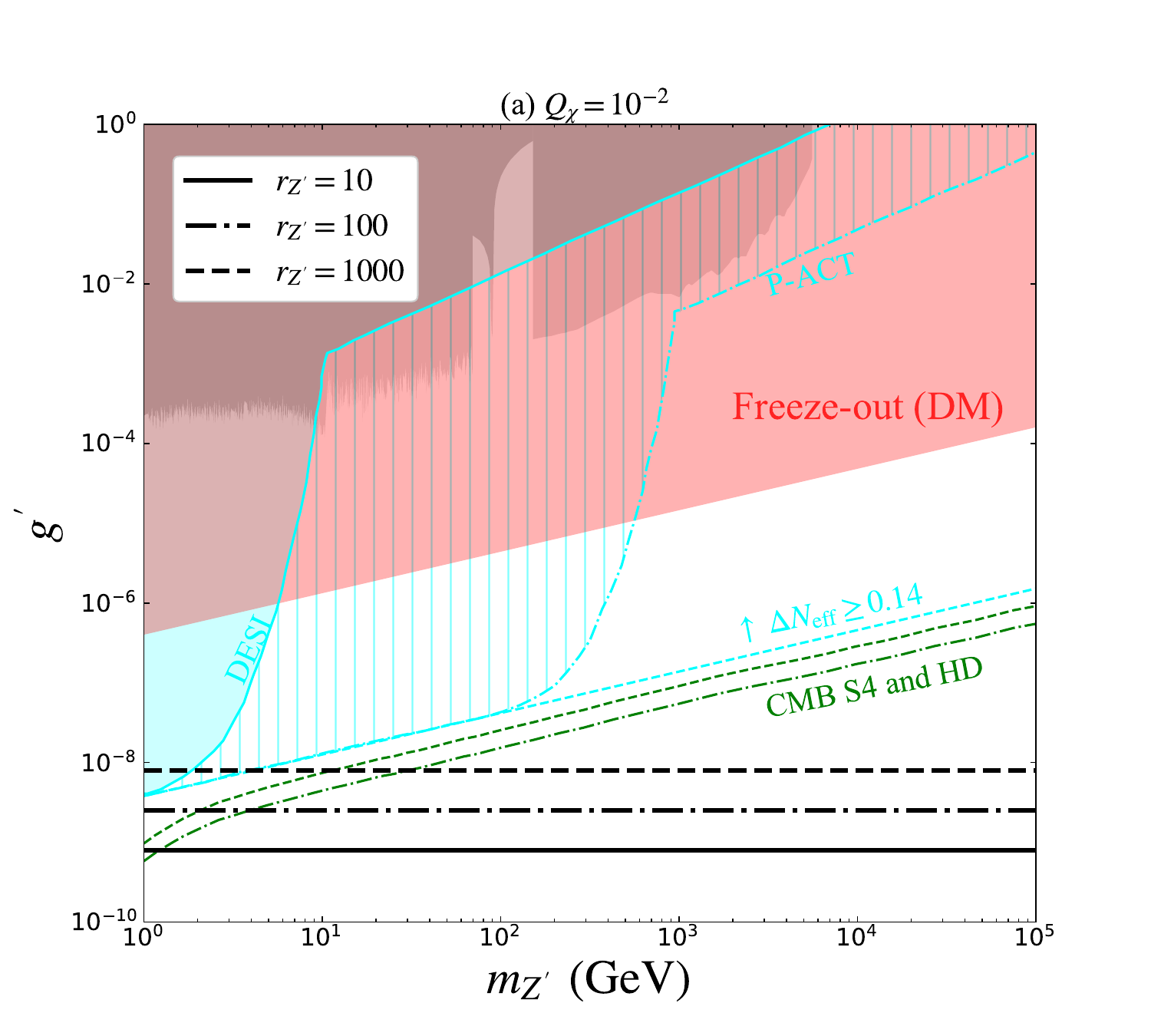}
  		\includegraphics[width=0.45\linewidth]{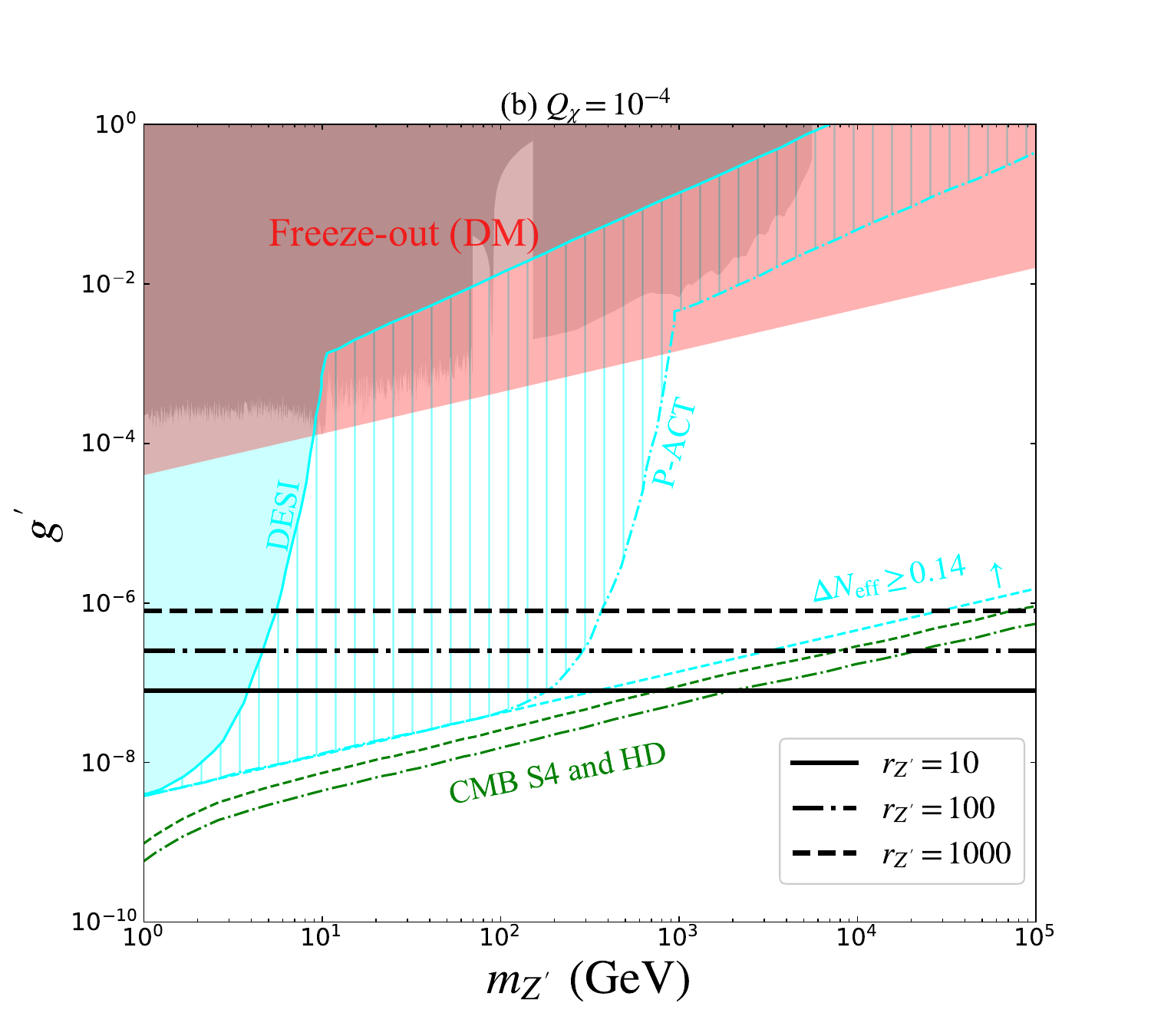}
  		\includegraphics[width=0.45\linewidth]{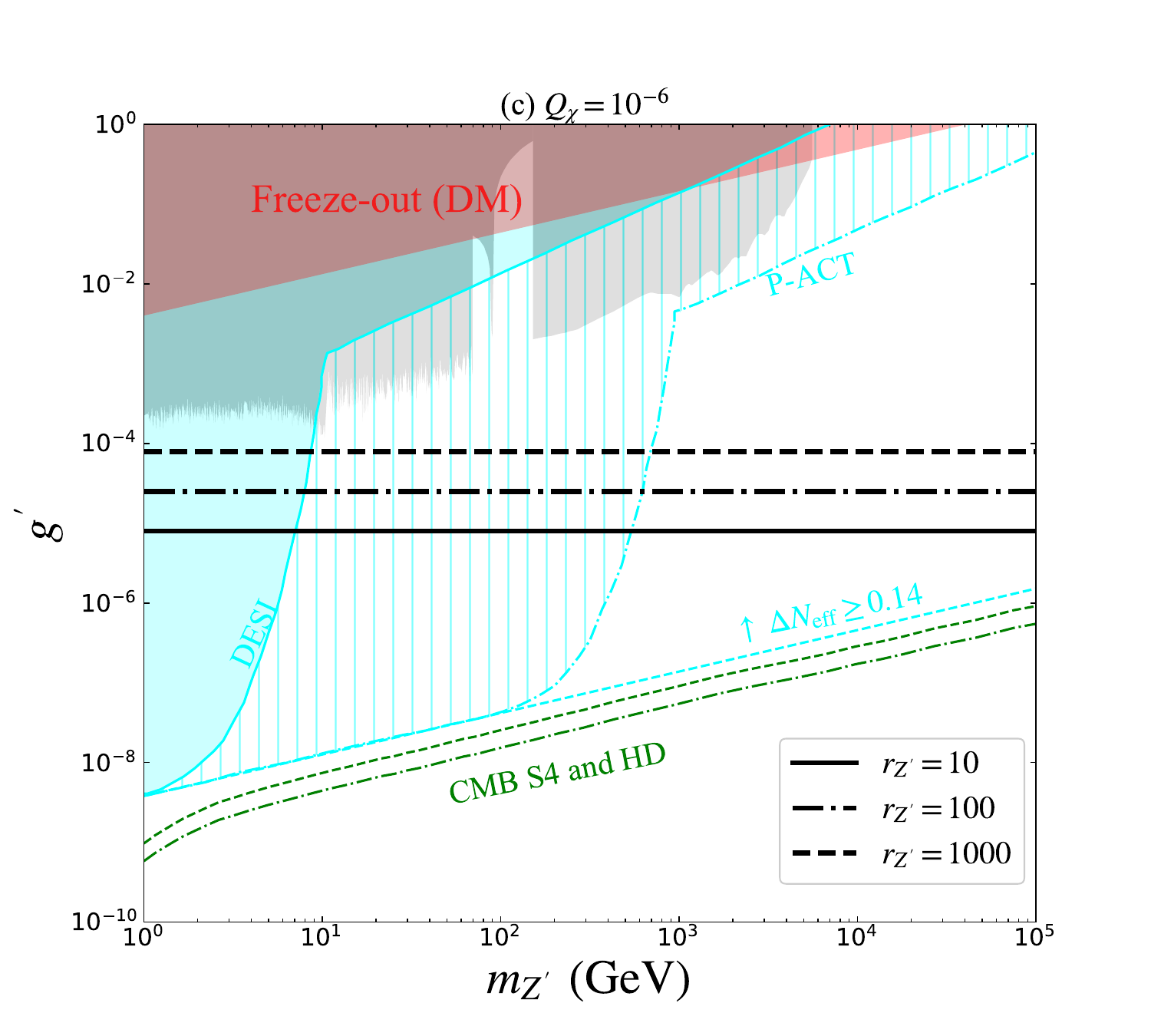}
  		\includegraphics[width=0.45\linewidth]{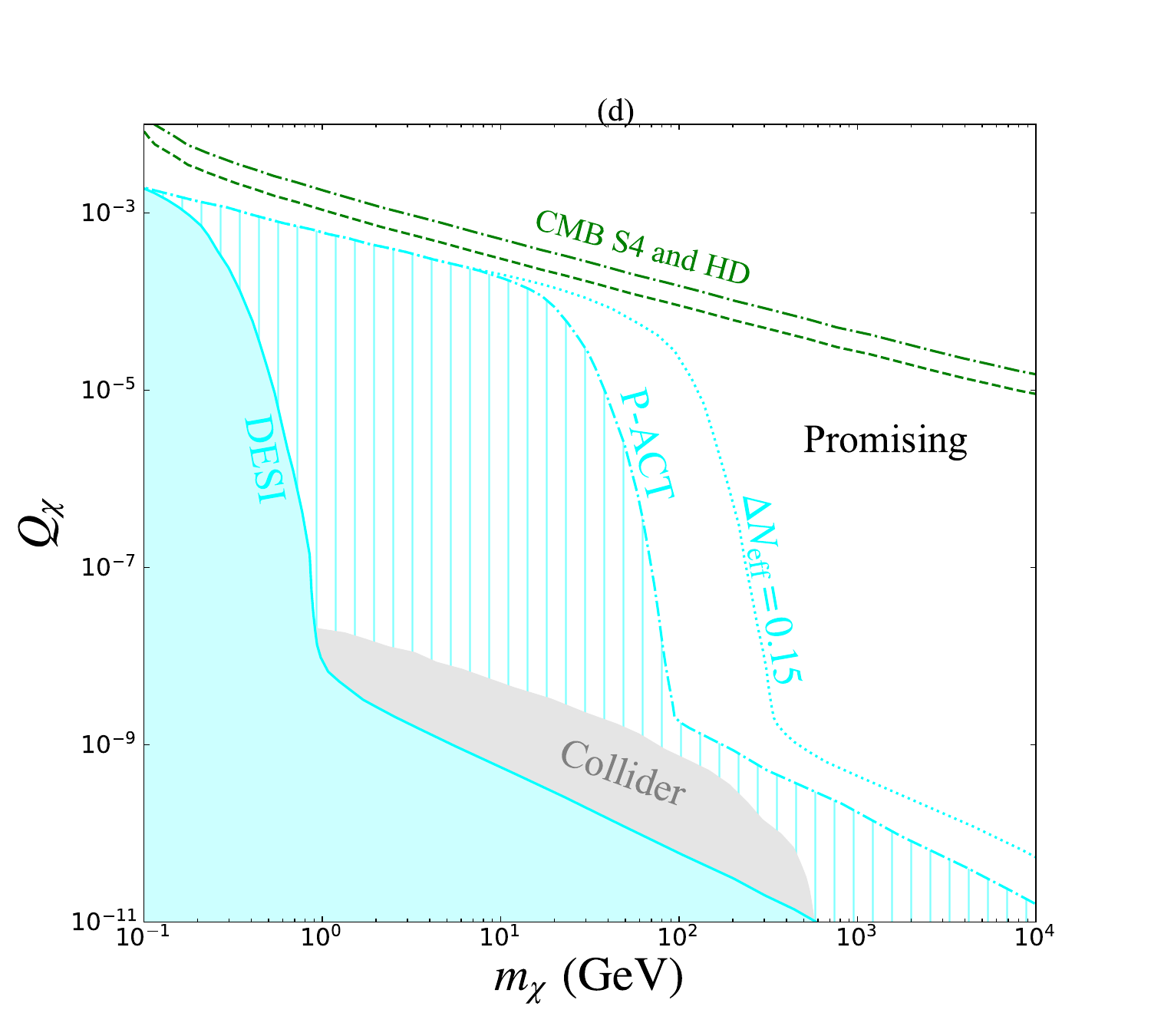}
  	\end{center}
  	\caption{Same as Figure~\ref{FIG:fig4}, but for the FIMP scenario. Panel (a), (b), and (c) correspond to cases of $Q_\chi=10^{-2}$, $Q_\chi=10^{-4}$, and $Q_\chi=10^{-6}$, respectively. In each case, the black solid, dot-dashed, and dashed lines represent the benchmark $r_{Z^\prime}=10$, 100, and 1000, respectively. Dark matter reaches thermal equilibrium in the red shadow, where is not considered in this FIMP scenario. The promising region in panel (d) is obtained with $r_{Z^\prime}=10$.
  	}
  	\label{FIG:fig7}
  \end{figure}

As already shown in Figure \ref{FIG:fig6}, the non-thermal contribution of  $\Delta N_{\rm eff}$ might be too small to be detected in the future experiments when $g'\lesssim\mathcal{O}(10^{-9})$. On the other hand, the correct relic density requires the DM coupling $g_\chi=g'\times Q_\chi\sim\mathcal{O}(10^{-11})$ when $m_{Z'}=100 m_\chi$ according to Equation \ref{Eqn:rd}. Therefore, to produce an observable $\Delta N_{\rm eff}$, $Q_\chi\lesssim10^{-2}$ should be satisfied for FIMP DM. In panels (a), (b), and (c) of Figure~\ref{FIG:fig7}, we then select three representative benchmarks $Q_\chi=10^{-2}$, $10^{-4}$, and $10^{-6}$ to illustrate the collider and $\Delta N_{\rm eff}$ constraints on the correct dark matter.

Since the couplings of DM are extremely small, the resulting DM-nucleon scattering cross section and DM annihilation cross section are far below the sensitivity of direct detection and indirect detection experiments. Therefore, we omit the presentation of these two types of constraints in this FIMP scenario. The collider constraints  induced by the visible decays of $Z^\prime$ are consistent with those in the WIMP scenario.  As $Q_\chi$ decreases, the upper bound of the non‑thermal region for dark matter gradually rises, as indicated by the red region in the figures, which is determined by the DM coupling $g_\chi$.  For $m_{Z^\prime}$ increasing from 1 GeV to $10^5$ GeV, $g_\chi$ increases from $\mathcal{O}(10^{-9})$ to $\mathcal{O}(10^{-6})$. And it is almost independent of $r_{Z^\prime}$ with the condition $r_{Z^\prime}\ll1$ as in Equation \eqref{Eqn:gx}.   Unavoidably, $\nu_R$ is thermally produced in the non‑thermal dark matter regime with larger $g^\prime$, where the contribution  of non‑thermal dark matter  to $\nu_R$ does not need to be considered. Using the method from the WIMP scenario in Subsection \ref{W-rd}, we obtain the $\Delta N_{\rm eff}$ constraints from DESI, P-ACT, and thermalization. In the non‑thermal regime of $\nu_R$, we take into account the CMB-S4 and CMB-HD constraints from Ref.~\cite{Adshead:2022ovo}.

In panel (a) of Figure~\ref{FIG:fig7}, we show the benchmark scenario with $Q_\chi=10^{-2}$, where $\nu_R$ is mainly from the non-thermal production. The future CMB experiments could probe $m_{Z'}$ around $\mathcal{O}(10)$ GeV in this case. For the benchmark scenario with $Q_\chi=10^{-4}$, the non-thermal  dominant region is above the TeV-scale. And the current P-ACT limit has excluded $m_{Z'}\lesssim 10^2$ GeV in this scenario. When $Q_\chi$  becomes tiny, e.g., $Q_\chi=10^{-6}$, the contribution of $\Delta N_{\rm eff}$ is from the thermal production of $\nu_R$. Of course, if $\Delta N_{\rm eff}<0.14$ is confirmed in the future, such a case will be fully excluded. Based on Equation \eqref{Eqn:rd}, we report that increasing the mass ratio $r_{Z'}=m_{Z'}/m_\chi$ leads to a larger $g'$ for correct DM relic density with fixed $Q_\chi$, thus a larger value of $\Delta N_{\rm eff}$.

In panel (d) of Figure~\ref{FIG:fig7}, we take $r_{Z^\prime}=10$ as an example  to present the promising parameter space, which has a smaller $m_\chi$ as $r_{Z^\prime}$ increases.  When $Q_\chi\lesssim2\times10^{-3}$, a relatively small $m_\chi$ appears in the thermalization region of $\nu_R$, where $m_\chi$ increases as $Q_\chi$ decreases for the experimental limits. The P-ACT constraint induced by thermally produced $\nu_R$ determines the lower boundary of the promising region with a minimum $m_\chi=12.5$~GeV and maximum $Q_\chi=1.6\times10^{-4}$.  When $Q_\chi$  decreases to $\mathcal{O}(10^{-11})$, $m_\chi$ needs to be greater than $10^4$ GeV.  Even a smaller $Q_\chi$ has $g^\prime\gtrsim\mathcal{O}(1)$ for correct relic density as in Equation~\eqref{Eqn:rd}, which is excluded by the perturbation constraint. For different values of $\Delta N_{\rm eff}$, similar to the conclusion in the WIMP scenario, the stricter $\Delta N_{\rm eff}$ constraint has the smaller allowed region. Specifically, DESI corresponds to a minimum $m_\chi=1$ GeV and maximum $Q_\chi=2\times10^{-3}$, while $\Delta N_{\rm eff}\lesssim0.15$ requires $m_\chi\gtrsim50$ GeV and $Q_\chi\lesssim7\times10^{-5}$.  Additionally, when $Q_\chi\lesssim10^{-8}$, the corresponding $g^\prime\gtrsim10^{-4}$ for correct relic density, the collider constraints  exclude part of the parameter space within $1~\GeV\lesssim m_\chi\lesssim 550~\GeV$, which is already excluded by P-ACT.  In the non-thermal region of $\nu_R$, as $m_\chi$ increases from 0.1 GeV to $10^4$ GeV, the constraints of CMB-S4 and CMB-HD lie in the range of $\mathcal{O}(10^{-5})\lesssim Q_\chi\lesssim\mathcal{O}(10^{-3})$, which determines the upper bound of the detectable region. For $Q_\chi\gtrsim{O}(10^{-2})$ with corresponding $g'\lesssim\mathcal{O}(10^{-9})$, the predicted $\Delta N_{\rm eff}$ is beyond the scope of future experiments.

\section{Conclusion} \label{SEC:CL}

In the minimal $U(1)_{B-L}$ model containing dark matter $\chi$ and Dirac neutrinos $\nu_R$, the light Dirac neutrinos have considerable contribution  to the $\Delta N_{\rm eff}$, which is constrained by the current experiments DESI (requiring $\Delta N_{\eff}\lesssim0.4$) and P-ACT ($\Delta N_{\eff}\lesssim0.17$), as well as the future experiments CMB-S4 ($\Delta N_{\eff}\lesssim0.06$) and CMB-HD ($\Delta N_{\eff}\lesssim0.027$). Since there is a close relationship between the generation of dark matter $\chi$ and Dirac neutrino $\nu_R$ through $Z'$ portal, these $\Delta N_{\rm eff}$ constraints further restrict the allowed parameter space of dark matter. Building upon the dark matter detection constraints and the collider searches of $Z^\prime$, this model predicts promising dark matter regions that can be tested by the future  $\Delta N_{\rm eff}$ experiments. Based on the production mechanism of dark matter, detailed discussions are carried out in both the WIMP and the FIMP scenarios.

In the WIMP scenario, the observation of the dark matter relic density appears at the resonant ($m_{Z'}\simeq 2 m_\chi$) and secluded ($m_{Z'}<m_\chi$ ) positions. In the resonance scenario, we fix $r_{Z^\prime} = 2.001$ to avoid the fate of being excluded by indirect detection constraints with extreme enhanced annihilation cross section. Under various constraints including P-ACT, the surviving parameter space satisfies $400~\GeV\lesssim m_\chi\lesssim\ 1.54\times10^5$~GeV with $0.7\lesssim Q_\chi\lesssim3400$. Meanwhile the corresponding gauge coupling is $\mathcal{O}(10^{-4})\lesssim g^\prime\lesssim\mathcal{O}(10^{-1})$. Hence dark matter always lies in the thermal region of $\nu_R$, which indicates that the  allowed region has $\Delta N_{\rm eff}$ no less than 0.14. In the secluded scenario, we fix $r_{Z^\prime} = 0.5$ to illustrate the allowed region, which meets $20~\GeV\lesssim m_\chi\lesssim 4.9\times10^4$ GeV with $30\lesssim Q_\chi\lesssim\mathcal{O}(10^7)$. The corresponding gauge coupling extends to $\mathcal{O}(10^{-9})\lesssim g^\prime\lesssim\mathcal{O}(10^{-1})$.  When  $Q_\chi\gtrsim5\times10^6$, $\nu_R$ is non-thermally generated, resulting in  $\Delta N_{\rm eff}$  being less than 0.14. Future CMB-S4 and CMB-HD are expected to verify the finally obtained allowed parameter spaces in both cases.

In the FIMP scenario,  dark matter could transition from the non-thermal regime of $\nu_R$ to the thermal regime as $Q_\chi$ decreases. Eventually, the promising parameter space lies in the region enclosed by the CMB-HD and P-ACT constraints, which requires $1.6\times10^{-11}\lesssim Q_\chi\lesssim\mathcal{O}(10^{-3})$ in the range of $0.1~\GeV\lesssim m_\chi\lesssim 10^4$ GeV with $r_{Z^\prime} = 10$ and corresponding gauge coupling $\mathcal{O}(10^{-9})\lesssim g^\prime\lesssim\mathcal{O}(10^{-1})$. Among this region, $Q_\chi\lesssim2\times10^{-3}$ is mainly constrained by P-ACT, while CMB-HD at $\mathcal{O}(10^{-5})\lesssim Q_\chi\lesssim\mathcal{O}(10^{-3})$ provides a detectable upper limit.

In summary, dark matter in both the secluded and FIMP scenarios could be generated in the non‑thermal region of $\nu_R$, thereby yielding $\Delta N_{\rm eff}$ smaller than 0.14, which corresponds to the minimum $\Delta N_{\rm eff}$ contributed by thermal $\nu_R$. In contrast, dark matter in the resonance scenario never leaves the thermal region of $\nu_R$, resulting in a minimum $\Delta N_{\rm eff}$ larger than 0.14. If the future experiments CMB-S4 and CMB-HD do not observe a deviation  of $N_{\rm eff}$ compared to the SM, the resonance scenario will be completely excluded, while the secluded and FIMP scenarios still have surviving parameter space. Meanwhile, the secluded scenario can be further tested by the indirect detection experiments.

\section*{Acknowledgments}

This work is supported by the National Natural Science Foundation of China under Grant No. 12505112, Natural Science Foundation of Shandong Province under Grant No. ZR2024QA138 and ZR2026QC0016, State Key Laboratory of Dark Matter Physics.

\allowdisplaybreaks

\appendix

\section{The evolution of  decoupling temperature and $d \log g_s(T)/ (3~d \log T)$}  \label{SEC:AEDT}

\begin{figure}
	\begin{center}
		\includegraphics[width=0.45\linewidth]{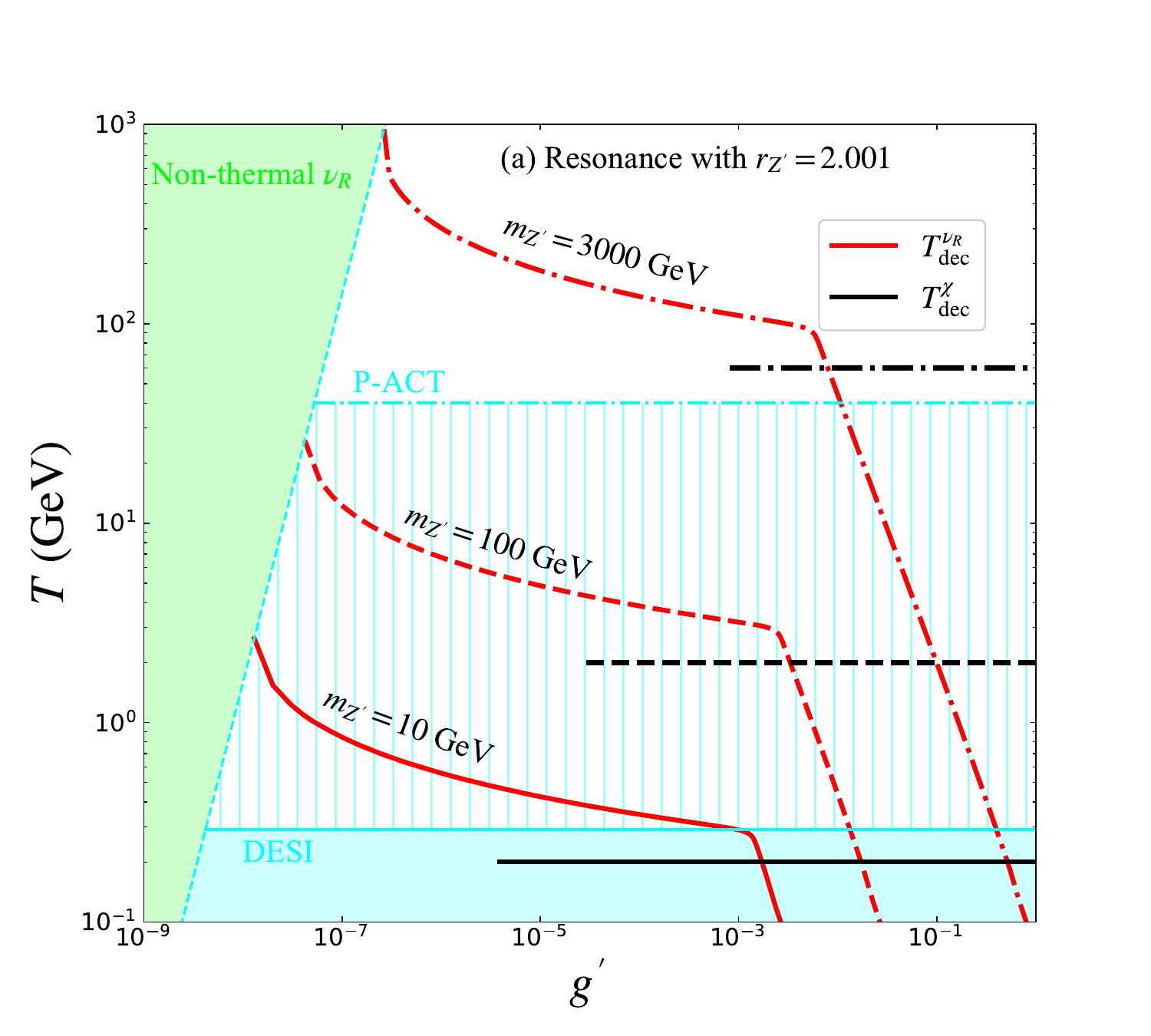}
		\includegraphics[width=0.45\linewidth]{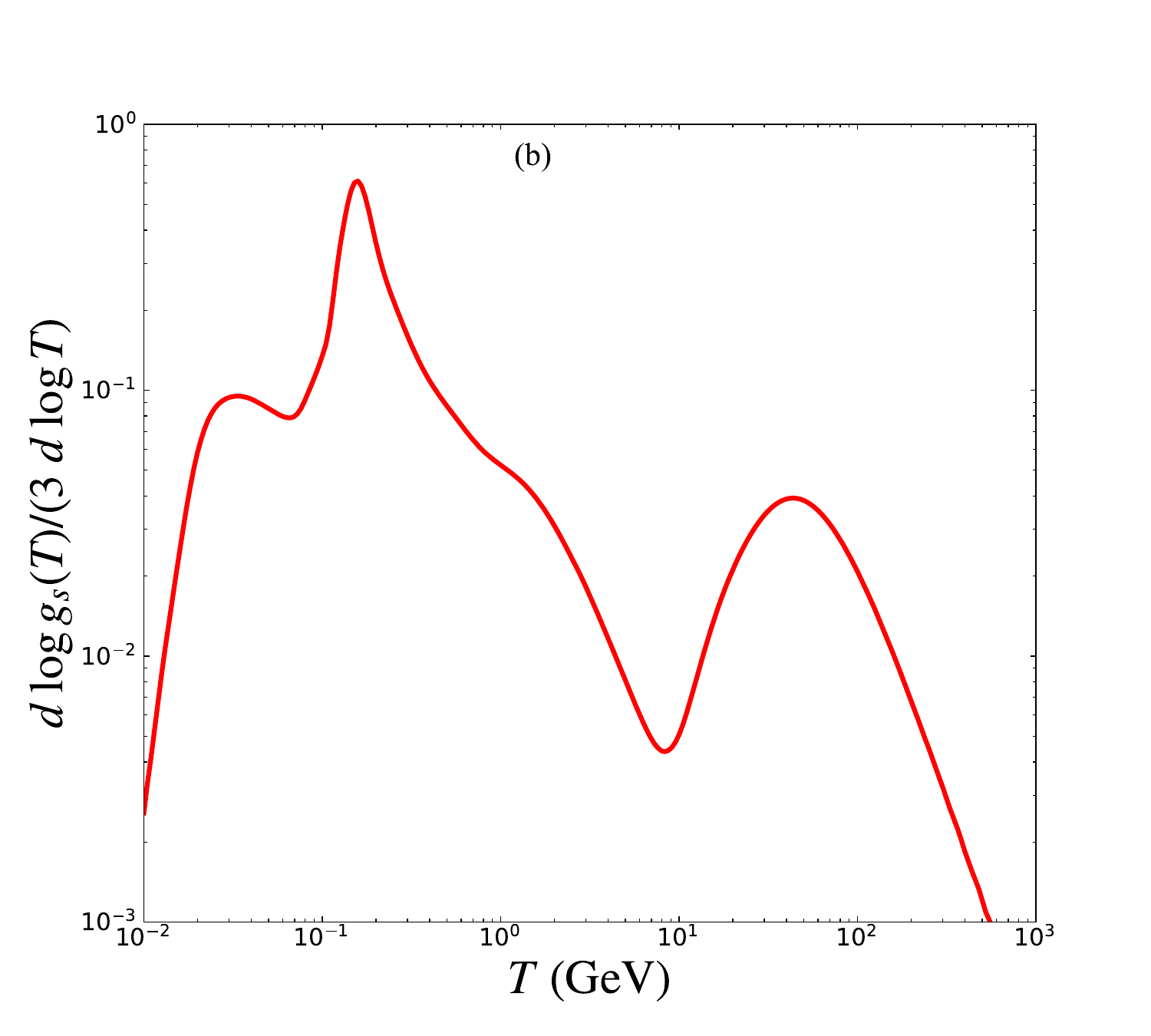}
	\end{center}
	\caption{The evolution of the decoupling temperature (a) in the resonance case and of the derivative of the SM dofs (b). In panel (a), the red and black curves represent $T_{\rm dec}^{\nu_R}$ and $T_{\rm dec}^{\chi}=m_\chi/25$, respectively. The solid, dashed, and dot-dashed lines correspond to $m_{Z^\prime} = $ 50 GeV, 100 GeV, and 3000 GeV, respectively. The cyan solid and dot-dashed lines represent the upper limits on $\Delta N_{\eff}$ given by DESI 2024 $\Delta N_{\eff}\simeq0.4$ \cite{DESI:2024mwx} and P-ACT $\Delta N_{\eff}\simeq0.17$ \cite{AtacamaCosmologyTelescope:2025blo,AtacamaCosmologyTelescope:2025nti}, respectively. The green region indicates that $\nu_R$ cannot reach thermal equilibrium.
	}
	\label{FIG:fig8}
\end{figure}

In the WIMP scenario, particularly under the resonance case, the $\nu_R\bar{\nu}_R\to \chi\bar{\chi}$ process exerts a considerable influence on the decoupling of $\nu_R$ when $T_{\rm dec}^{\chi}<T_{\rm dec}^{\nu_R}$. We show the range where this process is effective in Figure~\ref{FIG:fig8} (a). In this panel,  provided that the dark matter relic density is matched, different $m_{Z^\prime}$ implies different cutoff on $g^\prime$, and smaller $g^\prime$ would spoil the agreement as illustrated in Figure~\ref{FIG:fig4}.  As $m_{Z^\prime}$ increases from 10 GeV to 3000 GeV, the critical point where the $\nu_R\bar{\nu}_R\to \chi\bar{\chi}$ process ceases to be effective, i.e., $T_{\rm dec}^{\chi}=T_{\rm dec}^{\nu_R}$, corresponds to  $g^\prime$ rising from $1.7\times10^{-3}$ to $8\times10^{-3}$.  For $g^\prime$ below this critical value, $\nu_R\bar{\nu}_R\to \chi\bar{\chi}$ remains operative until the  non‑thermal $\nu_R$ regime is reached, and its upper limit on  $g^\prime$ grows from $1.3\times10^{-8}$ to $2.6\times10^{-7}$ with increasing $m_{Z^\prime}$. Under the combined constraints from DESI and P-ACT, the region with $T_{\rm dec}^{\chi}<T_{\rm dec}^{\nu_R}$ can usually  evade the limits for  TeV $m_{Z^\prime}$.  In the secluded scenario, $T_{\rm dec}^{\chi}$ still corresponds to $m_\chi/25$, but we do not consider the contribution of $\nu_R\bar{\nu}_R\to \chi\bar{\chi}$ which is suppressed by off-shell ${Z^\prime}$. Therefore, regardless of the relative magnitudes of $T_{\rm dec}^{\nu_R}$ and $T_{\rm dec}^{\chi}$,  $\nu_R\bar{\nu}_R\to f\bar{f}$ always dominates the decoupling of $\nu_R$. We do not show the corresponding figure here.

In panel (b) of Figure~\ref{FIG:fig8}, we show the evolution of $d \log g_s(T)/ (3~d \log T)$  within the decoupling temperature range of dark matter and neutrinos of our primary interest. This factor  appears as an additive term in the Boltzmann equations and is compared with  1. Even near the QCD phase transition $\sim160$ MeV, the maximum value of this factor remains below 1. We therefore neglect its contribution in the Boltzmann equations.

\section{The cross sections of various processes}  \label{SEC:ACS} 

The resonance annihilation cross sections of dark matter are given by

\begin{eqnarray}
	\sigma (s^\prime)_{\chi \bar{\chi} \to  f \bar{f}} = \frac{N_c^{f} Q_\chi^2 Q_f^2 {g^\prime}^4}{12 \pi s^\prime} \frac{\sqrt{s^\prime - 4 m_{f}^2}}{\sqrt{s^\prime - 4 m_\chi^2}} \frac{\left( s^\prime + 2 m_\chi^2\right) \left( s^\prime + 2 m_{f}^2 \right)}{ \left[ (s^\prime- m_{Z^\prime}^2)^2 + m_{Z^\prime}^2 \Gamma_{Z^\prime}^2 \right]},
\end{eqnarray}
where the total decay width of $Z^\prime$ satisfies $\Gamma_{Z^\prime}=\Gamma_{Z^\prime\to \chi\bar{\chi}}+\Gamma_{Z^\prime\to f\bar{f}}$ with
\begin{eqnarray}\label{Eqn:zpxx}
	\Gamma_{Z^\prime\to \chi\bar{\chi}}&=&\frac{{g^\prime}^2 Q_\chi^2 ~m_{Z^\prime}}{12\pi}\left(1+\frac{2 m_{\chi}^2}{m_{Z^\prime}^2}\right)\sqrt{1-\frac{4 m_{\chi}^2}{m_{Z^\prime}^2}},\\
	\Gamma_{Z^\prime\to f\bar{f}}& = &\sum_f \frac{N_c^f {g^\prime}^2 Q_f^2 m_{Z^\prime}}{12\pi}\left(1+\frac{2 m_f^2}{m_{Z^\prime}^2}\right)\sqrt{1-\frac{4 m_f^2}{m_{Z^\prime}^2}}, 
\end{eqnarray}
where $N_c^f$ is the color number of $f$.

For the secluded process,
\begin{eqnarray}
	\sigma (s^\prime)_{\chi \bar{\chi} \to  Z^\prime Z^\prime}&=&\frac{Q_\chi^4{g^\prime}^4}{8 \pi s^\prime} \frac{a}{b} \bigg[ -1 -\frac{(2+c^2)^2}{(2-c^2)^2-4 b^2}+\frac{6-2c^2+c^4+12b^2+4b^4}{2ba(1+b^2+ a^2)}\\ \nonumber
	&\times& \log \left(\frac{1+(b+a)^2}{1+(b-a)^2}\right) \bigg],
\end{eqnarray}
with $a=\sqrt{\frac{s^\prime}{4m_\chi^2}-\frac{m_{Z^\prime}^2}{m_\chi^2}}$, $b=\sqrt{\frac{s^\prime}{4m_\chi^2}-1}$ and $c=\frac{m_{Z^\prime}}{m_\chi}$.

For the  process of  $\nu_R \bar{\nu}_R \to f\bar{f}$,
\begin{eqnarray}
	\sigma (s^\prime)_{\nu_R \bar{\nu}_R \to f\bar{f}}= \frac{N_c^{f}Q_{\nu_R}^2 Q_{f}^2 {g^\prime}^4}{12\pi\sqrt{s^\prime}}  \frac{( s^\prime + 2 m_f^2 ) \sqrt{s^\prime -4 m_f^2 }}{  (s^\prime- m_{Z^\prime}^2) + m_{Z^\prime}^2 \Gamma_{Z^\prime}^2 }.
\end{eqnarray}

\section{The calculation of $\Delta N_{\eff}$}  \label{SEC:ANEFF}

The effective number of the relativistic degrees of freedom $\Delta N_{\eff}$ from three generations of $\nu_R$ can be defined as
\begin{eqnarray}\label{Eqn:dneff}
\Delta N_{\eff}=3 \times\frac{8}{7}\times\left(\frac{11}{4}\right)^{4/3}\times\frac{\rho_{\nu_R}}{\rho_\gamma}
\end{eqnarray}
In the thermal regime, when $\nu_R$ decouples from the SM plasma at $T_\text{dec}^{\nu_R}$, this equation can be simplified to 
\begin{eqnarray}
	\Delta N_{\eff}&=&3\times\left(\frac{11}{4}\right)^{4/3}\times\left(\frac{T_\text{dec}^{\nu_R}}{T_\gamma}\right)^4 \\ \nonumber
	&=& 3\times\left(\frac{11}{4}\right)^{4/3} \times\left(\frac{g_\star({T_\gamma})}{g_\star (T_\text{dec}^{\nu_R})}\right)^{4/3}\\
	&\simeq& 3\times0.047\times\left(\frac{106.75}{g_\star (T_\text{dec}^{\nu_R})}\right)^{4/3}
\end{eqnarray}
with $\rho_\gamma=\pi^2 T_\gamma^4/15$, $\rho_{\nu_R}=7\pi^2 ({T_\text{dec}^{\nu_R}})^4/120$ and $g_\star({T_\gamma})=43/11$.

In the non‑thermal regime, $\nu_R$ fails to attain thermal equilibrium. Therefore, we do not expand $\rho_{\nu_R}$ with respect to temperature. By applying the relation 
\begin{eqnarray}
\rho_{\nu_L}=\frac{7}{4}\times \frac{\pi^2}{30}\times T_{\nu_L}^4=\frac{7}{4}\times \frac{\pi^2}{30}\times T_{\gamma}^4\times\left(\frac{4}{11}\right)^{4/3}=\frac{7}{8}\times\left(\frac{4}{11}\right)^{4/3}\times\rho_{\gamma}
\end{eqnarray}
where we use $T_{\nu_L}/T_{\gamma}=(4/11)^{1/3}$ at CMB epoch. We derive Equation~\eqref{Eqn:dneff} anew as
\begin{eqnarray}
\Delta N_{\rm eff}=2\times3\times\left(\frac{\rho_{\nu_R}}{\rho_{\nu_L}}\right)_{\rm CMB}=6 \times\left(\frac{s^{4/3}Y_{\nu_R}}{\rho_{\nu_L}}\right)_{T=10~\MeV}
\end{eqnarray}
where the factor 2 accounts for $\nu_R$ and $\bar{\nu}_R$, and  is absorbed into the expansion of $\rho_{\nu_R}$ in the thermal scenario.

	
\end{document}